\documentclass{ws-rv9x6}
\usepackage{subfigure}   
\usepackage{ws-rv-thm}   
\usepackage{ws-rv-van}   
\usepackage{hyperref}
\makeindex

\begin{document}


\author{Yujun Wang}
\address{Department of Physics, Kansas State University, Manhattan, Kansas, 66506, USA}
\author{P. S. Julienne}
\address{Joint Quantum Institute, University of Maryland and NIST, College Park, Maryland 20742, USA}
\author{Chris H. Greene}
\address{Department of Physics, Purdue University, West Lafayette, Indiana 47907-2036, USA}

%

\body
\setcounter{chapter}{1}
\chapter{Few-body physics of ultracold atoms and molecules with long-range interactions\label{Fewbody}}

\section{Introduction}\index{Intro} 

The quantum mechanical few-body problem at ultracold energies poses severe challenges to theoretical techniques, particularly when long-range interactions are present that decay only as a power-law potential.  One familiar result is the modification of the elastic scattering Wigner threshold laws \cite{sadeghpour2000jpb} 
in the presence of power-law interaction potentials. Analogously, the near-threshold behaviors of all two-body scattering and bound state observables require generalizations in order to correctly describe the role of long range potentials.  
These generalizations are often termed modified effective range theories \cite{omalley1961jmp,idziaszek2006pra,friedrich2009pra,friedrich2011pra} 
or generalized quantum defect theories \cite{greene1979prab, greene1982prab, watanabe1980pra, mies1984jcpa, mies2000prab, burke1998prlb, gao1998pra,gao1998bpra, gao2001pra, gao2005pra, ruzic2013pra, idziaszek2011njp, gao2013pra, ruzic2013pra}  
in the literature.  

For simple, isotropic long-range interaction potentials involving one or more power-law potentials, analytical solutions are often known at least at zero energy, and allow extensive development in terms of closed-form solutions through the techniques of classical mathematics. \cite{seaton1983rpp, watanabe1980pra, gao1998bpra}
  
With anisotropic long-range interactions however, such as the dipole-dipole interaction characteristic for a gas composed of polar molecules or else of strong magnetic dipoles, often the theoretical heavy lifting relies on numerical solution techniques. \cite{ ni2010nature, quemener2012cr, blume2012rpp}
For dipolar scattering, numerical methods provide critical detailed information beyond the insights gleaned from simpler analytical approaches such as the partial-wave Born approximation. \cite{kanjilal2008pra, bohn2009njp, cavagnero2009njp, wang2012pra, dincao2011pra, ticknor2011pra, giannakeas2013prl}

Moving from the problem of two interacting particles to three or more involves a tremendous leap in the complexity of the theoretical description, as well as a commensurate richness in the phenomena that can occur.  For three interacting particles having no long range Coulomb interactions, the realm of universal Efimov physics has received tremendous theoretical interest \cite{amado1972prd, amado1973prd, efimov1972jetpl, efimov1973npa, efimov1979sjnp, macek1986zpd, adhikari1988pra, esry1996jpb, esry1996pra, nielsen1999prlb, esry1999prl, bedaque2000prl, nielsen2001prep, macek2006pra, deltuva2010prab, deltuva2010pra, efremov2009pra, wang2010prl, wang2011njp, wang2012prl, naidon2014prl,wang2014natphys} during the several decades since its original prediction\cite{efimov1970plb, efimov1971sjnp} in 1970. 
Moreover, it has received extensive experimental attention since the first clear observation of the Efimov effect in the 2006 measurement by the Innsbruck group of Grimm and coworkers.\cite{kraemer2006nature} 
The community of theorists interested in the Efimov effect started initially in nuclear physics, but the recent studies have concentrated on ultracold atomic and molecular systems.  This change in the relevant subfield occurred because the key parameter controlling Efimov states is the atom-atom scattering length, which is nowadays controllable near a Fano-Feshbach resonance,\cite{fano1935nc, feshbach1958ap, fano1961pr, fano2005J.Res.Natl.Inst.Stand.Technol., feshbach1962ap} through the application of external fields.\cite{stwalley1976prl, tiesinga1993pra, inouye1998NT, chin2010rmp}
 
One of the interesting aspects of the Efimov effect is that for sufficiently large two-body scattering lengths, the sum of pairwise particle interactions produces a net effective attraction, namely a hyperradial potential energy curve proportional to ${-1/R^2}$ where $R$ is the three-body hyperadius [see Eq.~(\ref{Eq_Hyperradius})].  
Such a potential is sometimes called a ``dipole potential'' because it also 
arises in various atomic and molecular two-body systems, in particular when a charge moves in the field of a permanent dipole, which is either a polar molecule\cite{garrett2970cpl, engelking1984pra, lineberger1997pra, clark1980jpb} or else an excited, degenerate hydrogen atom \cite{gailitis1963procphyssoc, seaton1961procphyssoc, greene1980jpb}.  A peculiarity of such an attractive inverse square potential is that it is too singular to produce a unique solution near the origin, and that behavior must be regularized in any given physical system.  For the charge-dipole system, it occurs naturally because any dipole has finite extent and so the dipole potential does not hold once the charge moves inside the charge distribution that produces the dipole.  In Efimov physics, the short-range hyperradial phase as the system moves into the effective dipole potential (Efimov) region of the hyperradius is regulated by the so-called ``three-body parameter''. \cite{braaten2006prep, dincao2009jpb} 

We discuss the details of the three-body 
parameter in Sec.~\ref{3BP}. The main idea is the following: 
in the presence of zero-range two-body interactions, the energy of the lowest Efimov resonance would drop to negative infinity were it not for the fact that the $-1/R^2$ 
potential energy does not hold all the way down to $R=0$ 
for any real system.  Zero-range interactions are a convenient model of Efimov physics, 
but they require the introduction of a three-body 
parameter to truncate the ground state energy to a finite value. One can view the three-body parameter as setting either a characteristic length or energy related to the deepest Efimov trimer.  When expressed as an energy, it is the lowest Efimov energy level at unitarity, and when expressed as a length it is the value of the negative scattering length $a_{-}^*$ at which the lowest Efimov state binding energy goes to zero (for $a<0$). 
In some models, such as effective field theory \cite{braaten2006prep} 
or in a model analysis of the adiabatic hyperspherical potential curves, \cite{wang2013amop,  mehta2009prl} it is possible to derive an approximate relationship between these two different ways of defining the three-body parameter.  In the first few decades of theoretical study of Efimov physics, it was taken as a matter of fact that the three-body parameter could not be predicted in general on the basis of two-body interactions alone, and that for different real systems that parameter would vary almost randomly.  This view was expressed, for instance, in \cite{dincao2009jpb}.  Thus it was a major surprise when experiments on the system of three 
bosonic Cs atoms began to show\cite{berninger2011prl} that several different Efimov resonances have almost the same value of the three body parameter expressed as $a_-^*$. 
In particular, expressed in units of the van der Waals length $r_{\rm vdW}$, 
which is defined by \cite{gribakin1993pra,chin2010rmp} 
\begin{equation}
r_{\rm vdW}=\frac12 (2\mu_2 C_6/\hbar^2)^{1/4}
\end{equation} 
for two atoms of reduced mass $\mu_2$ interacting via the $-C_6/r^6$ van der Waals potential ($r$ is the two-body distance), 
experiments showed that $a_{-}^* / r_{\rm vdW} \approx -10$.  
Around one year later, theory was able to explain 
this quasi-universality of systems having a long-range van der Waals interaction between each pair of particles, but this eventually was explained 
for a system of three identical bosonic atoms \cite{wang2012prl} which was  
largely confirmed later by \cite{naidon2014prl} 
and for a three-boson system with only two 
of the atoms identical by \cite{wang2012prlb}.

It remains of considerable interest to map out this recently identified quasi-universality of the three-body parameter to see how it depends on the nature of the long-range two-body interactions.  Evidence that there could be a universality for interacting dipoles was in fact suggested in two studies of the three-dipole problem, 
first for three 
oriented bosonic dipoles where an Efimov effect was predicted for the first time in a 3D system where angular momentum is not conserved \cite{wang2011prl}.  That study was later extended to three 
oriented identical fermionic dipoles where there is no predicted Efimov effect, but nevertheless an interesting universal state could be predicted. \cite{wang2011prlb}

The present review summarizes recent developments that have led to an improved understanding of long-range interactions, focusing on two-body and three-body systems with ultracold atoms or molecules. While there have been tremendous accomplishments in experimental few-body physics in recent years, the present review concentrates on the theoretical understanding that has emerged from combined experimental and theoretical efforts. The theory has been developed to the point where we increasingly understand which aspects of few-body collisions, bound states, and resonances are universal and controlled only by the long-range Hamiltonian, and which aspects differ and distinguish one species from another having identical long-range forces.

\section{van der Waals physics for two atoms}\label{vdW_2atoms}\index{index!multiple indexes}

\subsection{van der Waals universality and Feshbach resonances with atoms} 
\label{vdW_2atoms} 

Let us consider two atoms interacting by a long range potential of the form $-C_N/r^N$, where $r$ is the distance between the atoms. 
While the physics associated with various values of $N$ 
has been widely studied, {\it e. g.}, $N=1$ (the Coulomb potential), $N=3$ (two dipoles, see Sec.~\ref{2dipoles}), or $N=4$ (an ion and an atom, see Sec.~\ref{LongRange}), in this section 
we consider the specific case of $n=6$ 
for the van der Waals (vdW) potential between two neutral $S$-state atoms.  
When the length $r'=r/r_{\rm vdW}$ and potential $v(r)=V(r)/E_{\rm vdW}$ 
are scaled by their respective van der Waals units of length and energy $E_{\rm vdW}=\hbar^2/(2\mu_2 r_{\rm vdW}^2)$, 
the long range potential $v(r')$ between the two atoms including the centrifugal barrier for partial wave $\ell$ is~\cite{jones2006rmp} 
\begin{equation}
  v(r') \to -\frac{16}{r'^6} +\frac{\ell (\ell+1)}{r'^2} 
  \label{Eq:v-vdW}
\end{equation} 
It is important to note that there are other conventions in the literature defining a length and corresponding energy associated with the van der Waals potential. 
Gao~\cite{gao1998bpra} uses $\beta_6 = 2 r_{\rm vdW}$ and Gribakin and Flambaum~\cite{gribakin1993pra} introduced the mean scattering length $\bar{a} = [ 4 \pi / \Gamma(\frac14)^2 ] r_{\rm vdW} \approx 0.955 978 \ldots r_{\rm vdW}$~\cite{gribakin1993pra}. 
The latter is especially useful for giving a simpler form to theoretical expressions.

Gao~\cite{gao1998bpra} has worked out the analytic solutions for the bound and scattering states of this potential, which are especially relevant to two- and few-body physics with cold atoms, which to a large extent is governed by the states near the $E=0$ collision threshold, where $E$ represents energy.  These analytic solutions are parameterized in terms of the $s$-wave scattering length $a$~\cite{gao2004jpb}, a threshold property of the scattering phase shift $\eta(k) \to -ka$ for the $\ell=0$ 
partial wave as collision momentum $\hbar k = \sqrt{2\mu_2 E} \to 0$.  
Specifying $a$ specifies the bound ($E<0$) and scattering ($E \ge 0$)  solutions away from $E=0$ for all $\ell$~\cite{gao1998pra,gao2009pra,gao2001pra}. For example, if $a=\infty$, there is not only an $s$-wave bound state at $E=0$, but also an $E=0$ bound state for partial waves $\ell = 4, 8, 12 \ldots$; 
similarly, an $E=0$ $p$-wave bound state exists if $a=2\bar{a}$ and a $d$-wave bound state if $a=\bar{a}$.  

It is much more effective to use the analytic solutions to the van der Waals potential instead of the solutions to a zero-range pseudopotential to characterize near-threshold cold atom physics, since the former are accurate over a much wider range of energy near threshold.  
This is a consequence of the fact that $r_{\rm vdW}$ tends to be much larger than the range of strong chemical interactions that occur when $r \ll r_{\rm vdW}$.  Thus, the van der Waals potential spans a wide range of $r$ between the chemical and asymptotic regions, and its solutions accurately span a range of bound state and collision energies large compared to $E_{\rm vdW}$ and large compared to energy scales relevant to cold atom phenomena; see Refs.~\cite{jones2006rmp,chin2010rmp}.  For example, Chin {\it et al.}~\cite{chin2010rmp} give examples showing the near threshold spectrum of bound states for $s$-waves ($\ell=0$) 
and other partial waves ($\ell >0$) on a scale spanning hundreds of $E_{\rm vdW}$. These bound state energies in units of $E_{\rm vdW}$ are universal functions of $a/r_{\rm vdW}$, that is, species-specific parameters like the reduced mass and the magnitude of $C_6$ are scaled out by scaling by $r_{\rm vdW}$ and $E_{\rm vdW}$, and the scaled $a$ captures the effect of all short range physics. 

When the scattering length is large and positive, the energy of the last $s$-wave bound state of the potential is universally related to the scattering length by the simple relation,
\begin{equation}
 E_{-1}^\mathrm{U} \approx -\frac{\hbar^2}{2\mu_2 a^2} \,. 
 \label{Eq:EU}
\end{equation}
It is simple to make universal corrections to this expression due to the van der Waals potential.  Using reduced units of $\epsilon = E/\bar{E}$ for energy and $\alpha=a/\bar{a}$ length, Gribakin and Flambaum~\cite{gribakin1993pra} and Gao~\cite{gao2004jpb} have developed corrections to the universal binding energy $\epsilon_{-1}^\mathrm{U} =-1/\alpha^2$, valid when $1 \ll \alpha < \infty$: 
\begin{equation}
  \epsilon_{-1}^\mathrm{GF} =-\frac{1}{(\alpha-1)^2}, \,\,\,\,\,\,\,\,\, \epsilon_{-1}^\mathrm{Gao} =-\frac{1}{(\alpha-1)^2} \left ( 1 + \frac{g_1}{\alpha-1} + \frac{g_2}{(\alpha-1)^2} \right ) \,,
  \label{Eq:EGao}
\end{equation}
where $g_1 = \Gamma(\frac14)^4/6\pi^2 -2 \approx 0.9179$ and $g_2 = (5/4)g_1^2 -2 \approx -0.9468$.  Furthermore, the universal van der Waals effective range correction $r_e$ 
(defined in Section~\ref{MQDT})
to the scattering phase shift for $E>0$ is given by~\cite{gao1998bpra,flambaum1999pra}:
\begin{equation}
  \frac{r_e}{\bar{a}} = \frac{\Gamma(\frac14)^4}{6 \pi^2} \left (  1 - \frac{2}{\alpha} + \frac{2}{\alpha^2}  \right )
\end{equation}
Note that the effective range expansion breaks down near $\alpha \to 0$, or $|\alpha| \ll 1$.  

Fano-Feshbach resonances are possible when a closed channel bound state is tuned in the vicinity of an open channel threshold.  
A closed spin channel is one with separated atom energy (or threshold energy) $E_c >E>0$ 
that, in the case of alkali metal species with Zeeman spin structure, has a bound state with a different magnetic moment than the separated atoms of the threshold open channel, the difference being $\mu_{\rm dif}$. 
The ``bare'' closed channel bound state energy tunes with magnetic field $B$ as $\mu_{\rm dif} (B-B_c)$, 
whereas when the bare closed and open channels are mixed by interaction terms in the Hamiltonian, the scattering length takes on the following form near the resonance pole:
\begin{equation}
   a = a_\mathrm{bg} - a_\mathrm{bg} \frac{\Delta}{B-B_0}\,,
\end{equation}
where $a_\mathrm{bg}$ represents the ``'background'' or ``bare'' open channel scattering length in the absence of coupling to the closed channel bound state and 
$\Delta$ is the resonance ``width.''  The field position of the pole at $B_0$ where $E_{-1}^\mathrm{U}=0$ is normally shifted from $B_c$, as described in the next paragraph.
It is the ability to tune the scattering length to any value using resonance tuning that makes cold atoms such good probes of few-body physics. 

It turns out to be very useful to characterize the properties of various Feshbach resonance in terms of universal van der Waals parameters.  In fact, Chin {\it et al}~\cite{chin2010rmp} introduced a dimensionless parameter $s_\mathrm{res} = (a_\mathrm{bg}/\bar{a})( \Delta \mu_{\rm dif}/\bar{E})$  
to classify the strength of a 
resonance.  
Broad, open channel dominated resonances are those with $s_\mathrm{res} \gg 1$, and narrow, closed channel dominated ones are those with $s_\mathrm{res} \ll 1$.  The former tends to behave like single open channel states characterized by the scattering length alone.  The latter tend to exhibit their mixed closed/open channel character, and can not be fully characterized by their scattering length alone.  However, all isolated resonances take on the following simple threshold Breit-Wigner form for the phase shift~\cite{julienne2006intconfap,blackley2014pra}, here given in its near-threshold limiting form using the leading terms in an expansion in $\kappa=k\bar{a}$, valid when $\kappa \alpha_{\rm bg} \to 0$:
\begin{equation}
\eta(\epsilon,B)= \eta_{\rm bg}(\epsilon) -\tan^{-1}\left(\frac{\kappa s_\mathrm{res}}
{\epsilon-m_{\rm dif}(B-B_0)}\right) \,, 
\label{Eq:eta_res}
\end{equation}
where the reduced slope $m_\mathrm{dif} = \mu_\mathrm{dif}/\bar{E}$, 
and the pole is shifted from the ``bare'' crossing by $B_0-B_c= \Delta \alpha_{\rm bg} (1-\alpha_{\rm bg})/(1+(1-\alpha_{\rm bg})^2)$.  The universal van der Waals Eq.~(\ref{Eq:eta_res}) depends on the background scattering length $\alpha_{\rm bg}$, the ``pole strength'' $s_\mathrm{res}$ in the numerator of the resonance term, and 
the slope $m_\mathrm{dif}$. 
When collision energy is large enough that the condition $\kappa \alpha_{\rm bg} \ll 1$ is not satisfied, then the simple threshold forms for the numerator and shift term in Eq.~\ref{Eq:eta_res} need to be replaced by the more complex energy-dependent universal van der Waals functions of $\alpha_{\rm bg}$ as explained in Refs.~\cite{julienne2006intconfap,blackley2014pra}

Julienne and Hutson~\cite{julienne2014pra} give examples with Li atoms illustrating the universal binding energy formulas in Eqs.~(\ref{Eq:EU}-\ref{Eq:EGao}). The experimentally measured and coupled channels calculated binding energies for the broad 832G resonance of $^6$Li with $s_\mathrm{res} = 59$ agree well with the Gao expression $E_{-1}^\mathrm{Gao}$, but clear departures are seen from the conventional universal formula $E_{-1}^\mathrm{U}$.  On the other hand, the 
more closed channel dominant 738G resonance of $^7$Li with $s_\mathrm{res} = 0.54$ shows clear departures of calculated and measured binding energies from these simple formulas as the field is tuned away from the pole position.  Blackley {\it et al.}~\cite{blackley2014pra} test the effectiveness of the effective range expansion 
for $\eta(E,B)$ for finite energies near the resonance pole for several different cases.  While this expansion tends to be quite good near the poles of broad resonances, it tends to fail for narrow resonances, whereas the universal van der Waals expression of Eq.~(\ref{Eq:eta_res}), or its energy-dependent generalization, tends to be much more accurate. 

\begin{figure}[h]
\centerline{\includegraphics[width=10cm]{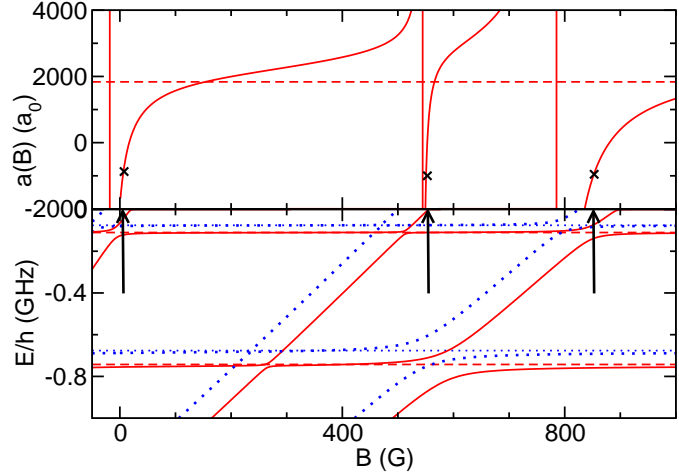}}
\caption{Scattering length (upper panel) and bound state energies (lower panel) versus magnetic field B for two interacting Cs atoms, each in its lowest Zeeman level with spin projection quantum number $+3$.  The solid lines show the $s$-wave results calculated using coupled channels calculations with an $s$-wave basis set only~\cite{berninger2013pra}; the heavy dotted lines show the corresponding $d$-wave levels with the same total spin projection (Ref.~\cite{berninger2013pra} also shows many other $d$ levels with other projection quantum numbers not shown here, as well as other resonances due to bound states of higher partial waves.)  The dashed lines in the lower panel show the  $-2$ and $-3$ universal van der Waals $s$-wave bound states corresponding to a ``background'' scattering length of $\alpha_\mathrm{bg}=18.3$ (dashed line in upper panel); the light dotted lines in the lower panel show the corresponding $d$-wave bound states.  The arrows and crosses indicate the regions where universal three-body Efimov states were observed~\cite{berninger2011prl}.  }
\label{fig1}
\end{figure}
Figure~\ref{fig1} illustrates some basic features of near-threshold bound states and the scattering length as $B$ is tuned for the case of two Cs atoms in their lowest energy Zeeman sub level, based on the theoretical coupled channels model in Ref.~\cite{berninger2013pra}.  There are three prominent broad $s$-wave resonances present with poles near 0 G, 550 G, and 800 G.  The corresponding ramping states are evident; the displacement between the apparent crossing of the linear ramping level from the nearby resonance pole position is a manifestation of the $B_0-B_c$ shift discussed above.  While the last $-1$ $s$-wave bound state is too close to threshold to be seen, the flat $-2$ and $-3$ bound states for a scattering length $\alpha_\mathrm{bg}=18.3$ are quite close to the actual calculated levels. Here the integer quantum number $-n$ counts the bound states down
from the threshold at $E=0$.
The agreement of calculated and universal van der Waals $s$- and $d$-wave levels illustrates how the pure van der Waals levels are good indicators of the levels of the real system, even on an energy scale large compared to $\bar{E}$ ($\bar{E}/h=$ 2.9 MHz in this case).  

It is also apparent that the different resonances in Fig.~\ref{fig1} overlap to some extent with one another. Reference~\cite{jachymski2013prl} shows how to extend the isolated resonance treatment above to a series of interacting resonances.   While it is still a good approximation to treat each resonance in an overlapping series as an isolated resonance near its pole, with a uniquely defined $s_\mathrm{res}$ parameter, its ``local'' background (in $B$ near a pole) can be a consequence of its interactions with other nearby resonances.  While it is usually adequate to consider resonances as isolated ones, we will give an example below in Sec.~\ref{ThreeSpinors} 
where the three-body Efimov physics of Cs atoms is affected by two overlapping resonances.

\subsection{Feshbach resonances and multichannel quantum defect theory (MQDT)} 
\label{MQDT}

Theories that are typically grouped under the name ``quantum defect theory'' can be viewed as having two main elements: (1) They systematically separate energy dependences arising from long-range interactions from those associated with short-range physics;  and (2) They connect and interrelate the physics of high-lying bound states (spectroscopy) with the properties of low-energy scattering states (collisions).  Historically, the first study along these lines was the development of effective range theory of Schwinger, Bethe, Fermi and Marshall, and others, which expresses the cotangent of the low-energy $s$-wave 
scattering phase shift 
as $k \cot \delta_{\ell=0} \rightarrow -1/a + r_{e} k^2/2$ 
and relates $a$ to the energy of a high-lying bound state when the scattering length is large and positive [see Eq.~(\ref{Eq:EU})] 

The most systematic separation of short-range and long-range influences on single-channel collisions and spectroscopy in the context of atomic Rydberg states was developed by \cite{seaton1983rpp}, who described his theoretical framework using the term ``quantum defect theory'' (QDT). An earlier forerunner of these ideas was a derivation by \cite{Hartree1928} 
of the single channel Rydberg formula for any system like an alkali atom with a single electron moving outside of a closed shell ion core.  Generalizations exist for effective range theory, e.g. to include energy dependences associated with long range potentials due to polarization \cite{omalley1961jmp}  
or repulsive Coulomb forces relevant in nuclear physics \cite{Berger1965PR}.  One key difference between these generalized effective range theories and quantum defect theories is that the former typically concentrate on developing low-energy expansions of phase shift 
behavior, whereas quantum defect theory normally relies on exact solutions as a function of energy and has a wider range of applicability as a result.
Quantum defect theory was also generalized to treat attractive or repulsive charge-dipole potentials $\pm 1/r^2$ \cite{greene1979prab,greene1982prab} 
and polarization potentials that are attractive \cite{watanabe1980pra, idziaszek2011njp, gao2013pra} or repulsive \cite{watanabe1982pra}.  

Typically quantum defect theories have a wider energy range of validity than the simpler effective range theories, because the latter rely on Taylor expansions of the long-range field parameters in powers of the threshold energy whereas the former utilize exact (often analytically known) properties.  Subsequent generalizations to treat cold collisions began in the 1980s in the context of chemical physics \cite{mies1984jcpa,mies1984jcpb}.  
Later when ultracold collisions became crucial to understand for the burgeoning field of dilute quantum gases, theories aiming more concretely at describing two-body interactions with long range van der Waals tails were developed in earnest \cite{julienne1989josab,gao1998bpra, gao2001pra, burke1998prlb, mies2000prab, gao2005pra}, and they all demonstrated great efficiency in characterizing the energy dependence of single-channel and multichannel solutions to the coupled equations whose solutions govern the properties of Fano-Feshbach resonances.  One of the most recent generalizations and improvements to the theory by \cite{ruzic2013pra}
has been applied to the $^6$Li-$^{133}$Cs collision system \cite{Pires2014pra} 
where multichannel QDT (MQDT) predictions are compared with other treatments such as the full close-coupling solutions as well as the simpler and more approximate ``asymptotic bound state model'' (ABM) \cite{Tiecke2010pra}. 

The key idea of all these theories is that beyond some range $r>r_0$, it is usually an excellent approximation to treat the motion of separating particles as though they are moving in a simple, one-dimensional long-range potential $V_{lr}(r)$.  In a typical system involving cold collisions of ground state atoms, when particles separate beyond a distance of around $r_0 \approx 30 a.u.$, the potential energy can be accurately approximated by a van der Waals potential $-C_6/r^6$.  The solutions to the corresponding radial Schr\"odinger equation have been worked out in fully analytical form by \cite{gao1998bpra} 
but it is often simpler to rely on numerical QDT solutions to the radial differential equation.  Two linearly independent solutions in this theory are denoted $(f^{sr},g^{sr})$ where the superscript ``sr'' is meant to indicate that these are characterized by ``short range'' boundary conditions that guarantee these solutions are smooth and analytic functions of energy at small distances $r\le r_0$.

In the case of long range attractive single-power-law potentials $-C_N/r^N$ with $N\ge 3$, 
all linearly-independent solutions at small distance are oscillatory all the way down to $r\rightarrow 0$ even for nonzero centrifugal potentials $\ell(\ell+1)/2 \mu_2 r^2$ with $\ell>0$.  
Two energy-analytic solutions 
${\hat f, \hat g}$ in the notation of \cite{ruzic2013pra} 
can be identified in such cases by giving them equal amplitude near the origin and choosing a 90 degree phase difference. For instance, in the WKB approximation which is often adequate to specify boundary conditions in cold collision problems, but not essential, these analytic solutions have the following behavior in channel $i$ at small $r$: 

\begin{subequations}
\label{eq:fhatandghat}
\begin{align}
\label{eq:fhat}
\hat{f}_i(r)&=\frac{1}{\sqrt{k_i(r)}}\sin(\int_{r_\text{x}}^{r} k_i(r') \mathrm{d}r' + \phi_i) \quad\text{at }r=r_\text{x}\text{ },\\
\label{eq:ghat}
\hat{g}_i(r)&=-\frac{1}{\sqrt{k_i(r)}}\cos(\int_{r_\text{x}}^{r} k_i(r') \mathrm{d}r' + \phi_i) \quad\text{at }r=r_\text{x}\text{ },
\end{align}
\end{subequations}

where the point $r_\text{x}$ is some fixed small distance, and the phase $\phi_i$ is in general independent of energy and $r$ and is chosen according to some convenient criterion.  A choice particularly advantageous for $\phi_i$ at $\ell_i>0$ is the one suggested by \cite{ruzic2013pra},  
which picks the unique phase such that at zero channel energy, the exact zero-energy independent solution $\hat{g}_i(r)$ approaches $r^{-{\ell_i}}$ 
at $r \rightarrow \infty$ at channel energy $\epsilon_i = 0$.  Because the potential energy is deep in the region where the boundary conditions are chosen, far deeper than the energy range of interest in cold or ultracold collisions for practically any diatomic system, ${\hat f, \hat g}$ will be extremely smooth functions of energy over a broad range from well above a dissociation threshold to well below.  The usual set of real regular $f$ and irregular $g$ functions from scattering theory that oscillate 90 degrees out of phase at infinity with equal amplitudes are more convenient to use for defining physical reaction matrices. For any long range field one must find the energy- and $\ell$-dependent constant coefficients $A,{\cal G}$ relating ${f,g}$ to ${\hat{f}, \hat{g}}$, a relationship written as:  

\begin{subequations}
\label{eq:fhatandghat2}
\begin{align}
\label{eq:fhat2}
{f}_i(r)&=A_i^{1/2}\hat{f}_i(r),\\
\label{eq:ghat2}
{g}_i(r)&=A_i^{-1/2}({\cal G}_i \hat{f}_i(r)+\hat{g}_i(r)).
\end{align}
\end{subequations}

The physical meaning of these parameters $A,{\cal G}$ has been expounded in various references, although the notations and conventions are not always uniform among the various QDT references.  For instance, the long range QDT parameter $A^{1/2}$ can be viewed as containing the energy-rescaling factor needed such that the solution $f$ has the usual energy-normalized amplitude at $r\rightarrow \infty$.  As such, it contains the threshold law physics, i.e. the Wigner threshold laws and their generalizations appropriate to different long range potentials.  The parameter $A$ in this notation connects with the parameter $C(E)^{-1}$ in the notation of \cite{mies1984jcpa}, 
where also the long range parameter ${\cal G}$ is written as $\tan{\lambda(E)}$.  
The significance of the $C(E)$ and $\tan\lambda(E)$ functions for cold atomic collisions is discussed by~\cite{julienne1989josab,julienne2007acp} along with illustrative figures of their behavior.
The parameter $\lambda(E)$ is interpreted qualitatively as reflecting the fact that two linearly independent solutions that oscillate 90 degrees out of phase at small-$r$ generally cannot generally retain this phase difference all the way to $\infty$.  The asymptotic solutions of scattering theory (reaction matrices, etc.) require solutions 90 degrees out of phase at infinity, and relevant phase correction parameter $\lambda(E)$ or equivalently ${\cal G(E)}$ must be determined for each long range field of interest. 

A major difference between MQDT and ordinary scattering theory is that in multichannel quantum defect theory, one initially postpones enforcement of large-r boundary conditions in the closed channels, and works with an enlarged channel space that includes closed ($Q$) as well as open ($P$) channels. 
The linearly-independent short-range (sr) reaction matrix solutions in this enlarged channel space have the following form outside the radius $r_0$ beyond which the potential assumes its purely long-range form:

\begin{equation}
\Psi_{i'} = \sum_i \Phi_i (\hat{f}_i(r)\delta_{ii'}-\hat{g}_i(r) K^{\rm sr}_{ii'}). 
\end{equation}

These solutions are sometimes called ``unphysical'' because they diverge exponentially in the closed ($Q$) 
channels at any given energy, and to obtain the physical wavefunctions one must find superpositions of these linearly independent solutions that eliminate the exponentially growing terms.  This is straightforward once one writes the coefficient of the exponentially growing terms, namely

\begin{equation}
(\hat{f}_i(r)\delta_{ii'}-\hat{g}_i(r) K^{\rm sr}_{ii'})\rightarrow (\cos \gamma_i \delta_{ii'}+\sin \gamma_i K^{\rm sr}_{ii'}) \exp(\kappa_i r) 
\end{equation}

And the linear algebra that produces the physical reaction is simple, as has been derived in many other references, giving
\begin{equation}
K = K^{\rm sr}_{oo}- K^{\rm sr}_{oc} (K^{\rm sr}_{cc}+\cot \gamma)^{-1} K^{\rm sr}_{co} 
\end{equation}
In this form, the appearance of Fano-Feshbach resonances is manifestly clear through the matrix inversion that becomes singular near closed-channel resonances.  The smoothness of $K^{\rm sr}$ 
implies that it is often energy-independent and even field-independent to an excellent approximation over a broad range, i.e. over 0.1-1 K in energy and over hundreds of gauss.  Moreover the eigenvectors are expected on physical grounds to be approximately given by the recoupling transformation coefficients connecting short-range channels $|(i_1 i_2)I (s_1 s_2)S F M_F\rangle$ 
to the different coupling scheme appropriate at long range.  Specifically, in this short range coupling scheme, the nuclear spins $i_1,i_2$ and electronic spins $s_1,s_2$ of the individual atoms are first coupled 
to total nuclear and electronic spins $I$ and $S$ in the presence of 
strong electronic interaction, then couple to the total hyperfine angular momentum $F$ and its projection $m_F$ for the much weaker hyperfine interactions. 
At zero B-field, the asymptotic coupling that is relevant has good atomic hyperfine quantum numbers $|(i_1 s_1)f_1 m_1 (i_2 s_2) f_2 m_2\rangle$.  But when $B \ne 0$ one must diagonalize the Breit-Rabi Hamiltonian~\cite{breit1931pr,nafe1948pr}: 
\begin{equation}
\hat{h}_j= \zeta\hat{i}_j\cdot \hat{s}_j + g_e\mu_{\rm B}\boldsymbol{B}\cdot s_j + g_n \mu_{\rm B}\boldsymbol{B}\cdot i_j
\label{Eq_Breit-Rabi}
\end{equation}
 to find the atomic dissociation thresholds and eigenvectors  $\langle(i_1 s_1)f_1 | \lambda_1 \rangle^{m_1}$ and $\langle(i_2 s_2)f | \lambda_2 \rangle^{m_2}$ as well, in order to construct a first order approximation to the asymptotically correct channel states.  
In Eq.~(\ref{Eq_Breit-Rabi}) $g_e$, $g_n$ are electron and nuclear $g$ factors, $\mu_B$ is the Bohr magneton. 
The projection of these onto the short-range eigenchannels gives a unitary matrix ${\bold X}$ that plays the role of a frame transformation which can give an effective approximation scheme that does not require solution of any coupled equations.  Example 
applications of this multichannel quantum defect theory with frame transformation (MQDT-FT) are discussed by \cite{burke1998prlb,gao2005pra, Pires2014pra}.  Note that similar content is expressed in the ``three-parameter'' description of two-body Fano-Feshbach resonances as developed by \cite{Hanna2009pra}.

\subsection{Numerical predictions of Feshbach resonances} 

Although the MQDT has proved to be a very powerful tool to study low-energy scattering problems and to 
make quantitative predictions in many systems, in heavier atomic systems 
such as Cs, the importance of dipolar interaction~\cite{stoof1988prb,moerdijk1995pra}, second order spin-orbit coupling~\cite{chin2004pra,mies1996J.Res.Natl.Inst.Stand.Technol,leo1998prl,kotochigova2000pra}, and higher order dispersive potential ($-1/r^8$, $-1/r^{10}$, ...)~\cite{leo2000prl} 
make the Feshbach spectrum more complicated and difficult to be treated by the MQDT via frame transformation. It is therefore essential 
to numerically integrate the Schr{\"o}dinger equation to study the Feshbach physics in such systems. 

Numerical studies of atomic collisions relevant to ultracold experiments were firstly done for hydrogen atoms~\cite{Lagendijk1986prb,stoof1988prb}. Later studies 
then focused on alkali atoms~\cite{tiesinga1992pra,tiesinga1993pra,moerdijk1994prl,moerdijk1995pra,boesten1996pra,houbiers1998pra,burke1999pra,chin2004pra} 
with their cooling down to quantum degeneracy, and more recently on ultracold lanthanizes ---
Er and Dy~\cite{kotochigova2011pccp,petrov2012prl,frisch2014nature}. 
When including the spin degrees of freedom, ultracold atomic collisions can be solved as a standard coupled channel problem with 
typically a few to tens of channels. In 
the cases where the magnetic dipole-dipole interaction and/or the anisotropy of electronic interactions are not negligible, 
orbital angular momentum ceases to be a good quantum number and therefore needs to be coupled. 
This can lead to thousands of coupled channels in case of highly magnetic atoms like Dy or Er~\cite{kotochigova2011pccp,petrov2012prl,frisch2014nature,baumann2014pra} 
and therefore makes scattering studies 
quite numerically demanding, if scattering over a wide range of magnetic field is to be investigated. 

For collisions between alkali atoms that have been widely studied, only the lowest two Born-Oppenheimer potentials $^1\Sigma_g^+$ and $^3\Sigma_u^+$ are involved.  
Although these potentials are labeled by the total electronic spin $S$, it is more convenient to expand the scattering wave function in individual 
atom's electronic and nuclear spin basis where the Hamiltonian matrix can be efficiently evaluated (see, for example, Ref.~\cite{kohler2006rmp,chin2010rmp,berninger2013pra,hutson2008pra} for implementations). 
In the asymptotic region where interactions between atoms vanish, the scattering solution is projected to the eigenbasis of individual atoms --- the atomic 
hyperfine states (see Sec.~\ref{MQDT}) --- where the scattering matrix is extracted. 

The main challenge in the numerical study of ultracold collisions is the large integration range in the atomic separation, 
since the range should be at least a few times of the de Broglie wavelength. This typically means a range over $10^4$ Bohr 
and could be even larger when close to a Feshbach resonance. 
At small inter-atomic distance the Born-Oppenheimer potentials are generally deep enough for the scattering wave function to have hundreds of nodes, which 
require a good number of radial points or radial elements to represent. Although it is not a real issue for modern computers to solve such a problem 
in the simplest scenario, the calculations start becoming cumbersome when different partial waves are coupled by, for instance, the magnetic dipole-dipole 
interaction. 

To improve numerical efficiency, propagation methods are widely used in such scattering studies and also used in finding weakly-bound state 
energies of Feshbach molecules. For scattering calculations, a practical choice is the eigen-channel $R$-matrix propagation method~\cite{aymar1996rmp}, 
which is numerically stable and has the flexibility of being adaptive to different radial representations.

\section{van der Waals physics for three atoms}\label{vdW_3atoms}\index{index!multiple indexes} 

\subsection{Universal three-body parameter} 
\label{3BP}

The three-body parameter is central in studies of low-energy three-body physics. The origin of the parameter can be traced back to the early discovery of a 
peculiar quantum behavior for three particles --- the Thomas collapse~\cite{thomas1935pr}. In 1935, Thomas discovered that the ground state energy of three identical bosons 
diverges to $-\infty$ as the range of pairwise interactions $r_0$ shrinks to zero --- even if one keeps the two-body binding energy $E_{\rm 2b}$ 
a constant in this limiting process. 

The three-body parameter is essentially introduced to ``regularize'' the unphysical divergence of the three-body spectrum. 
In the effective field theory (EFT)~\cite{braaten2006prep} or in the momentum-space Faddeev equation~\cite{faddeev1961jetp}, 
the Thomas collapse is manifested as a logrithmic divergence of the three-body spectrum 
when the inter-particle momentum is taken to the ``ultraviolet'' limit~\cite{amado1971plb,amado1972prd,bedaque1999prl}. 
This divergence can be formally avoided by adding a three-body force 
that effectively introduces a cutoff in the inter-particle momentum, which assumes one form of the three-body parameter. 

The three-body parameter can be more intuitively understood in the hyperspherical coordinates~\cite{wang2010thesis}. In fact, this was the way used by Efimov to discover the 
Efimov effect~\cite{efimov1973npa}. The hyperspherical corrdinates are defined for many particles in analogy to the spherical coordinates for one particle, where only 
one of the cooridinates --- the hyperradius 
\begin{equation}
\label{Eq_Hyperradius} 
R=\sqrt{\frac{m_1 m_2 r_{12}^2+m_2 m_3 r_{23}^2+m_3 m_1 r_{31}^2}{\mu_3 (m_1+m_2+m_3)}}
\end{equation}
--- represents distance, whereas all other coordinates are defined as ``angles''. 
The three-body mass $\mu_3$ can be defined as 
\begin{equation}
\mu_3=\sqrt{\frac{m_1 m_2 m_3}{m_1+m_2+m_3}}.
\end{equation}
The advantage of the hyperspherical 
coordinates is primarily from $R$ being a universal breakup coordinate for all defragmentation processes. The hyperangles, represented by a collective of coordinates $\Omega$, 
can have different definitions according to the choice of orthogonal coordinates~\cite{wang2010thesis}. 

Fixing $E_{\rm 2b}$ in the limit $r_0\rightarrow 0$ is essentially equivalent to keeping the two-body scattering length $a$ unchanged. 
To study the behavior of three particles 
in the universal limit where $|a|\gg r_0$, 
it is easier to consider the special case with $|a|\rightarrow\infty$ and $r_0\rightarrow 0$. 
In this case it is easy to show~\cite{efimov1973npa} that the hyperadial motion separates from the hyperangular motions. 
The hyperangular motions can be solved analytically and 
lead to the following hyperradial equation: 
\begin{equation}
\label{Eq_Hyperradial_ZRP}
\frac{\hbar^2}{2\mu_3}\left(-\frac{d^2}{d R^2}+\frac{s_\nu^2-1/4}{ R^2}\right)F_\nu(R)=E_{3b} F_\nu(R), 
\end{equation}
where $s_\nu$ are universal constants~\cite{efimov1973npa}. 
For identical bosons and some combinations of mass-imbalanced three-body systems, the lowest $s_\nu^2$ is negative --- we denote it as $-s_0^2$ with $s_0\approx 1.00624$.
Eq.~(\ref{Eq_Hyperradial_ZRP}) then leads to the well-known ``fall-to-the-center'' problem~\cite{perelomov1970theomathphys} 
--- the three-body energy $E_{3b}$ has no lower bound. 
The problem comes, obviously, from the effective hyperradial potential 
\begin{equation}
U_\nu(R)=-\frac{s_0^2+1/4}{2\mu_3 R^2}\hbar^2
\end{equation}
being too singular near the origin. To avoid this three-body collapse, some regularization is needed such that the $-1/R^2$ potential does not extend all the way 
to $R=0$. A simple cure is to give a cutoff to the $-1/R^2$ potential, which is equivalent to giving an earlier mentioned momentum-space cutoff in the EFT. 
In either case, a cutoff leads to a finite three-body ground state energy $E_0$, or a binding wavenumber $\kappa^*=\sqrt{2\mu_3 E_0}/\hbar$ which is another 
form of the three-body parameter. 

Although the $r_0\rightarrow 0$ 
situation is not realized in nature, the unitarity condition where $|a|\rightarrow\infty$ is now routinely realized 
in ultracold atoms by using Feshbach resonances~\cite{chin2010rmp}. In this case, the $-1/R^2$ long-range potential leads to an infinite number of 
three-body bound states --- Efimov states --- with energies following a geometric scaling $E_{n+1}/E_n=e^{-2\pi/s_0}$, 
which can happen even when none of the two-body subsystems is really bound ($|a|\rightarrow\infty$ implies $E_{2b}\rightarrow 0$). This is 
known as the Efimov effect~\cite{efimov1970plb,efimov1973npa} in quantum physics. 

\begin{figure}
\centerline{\includegraphics[scale=0.3]{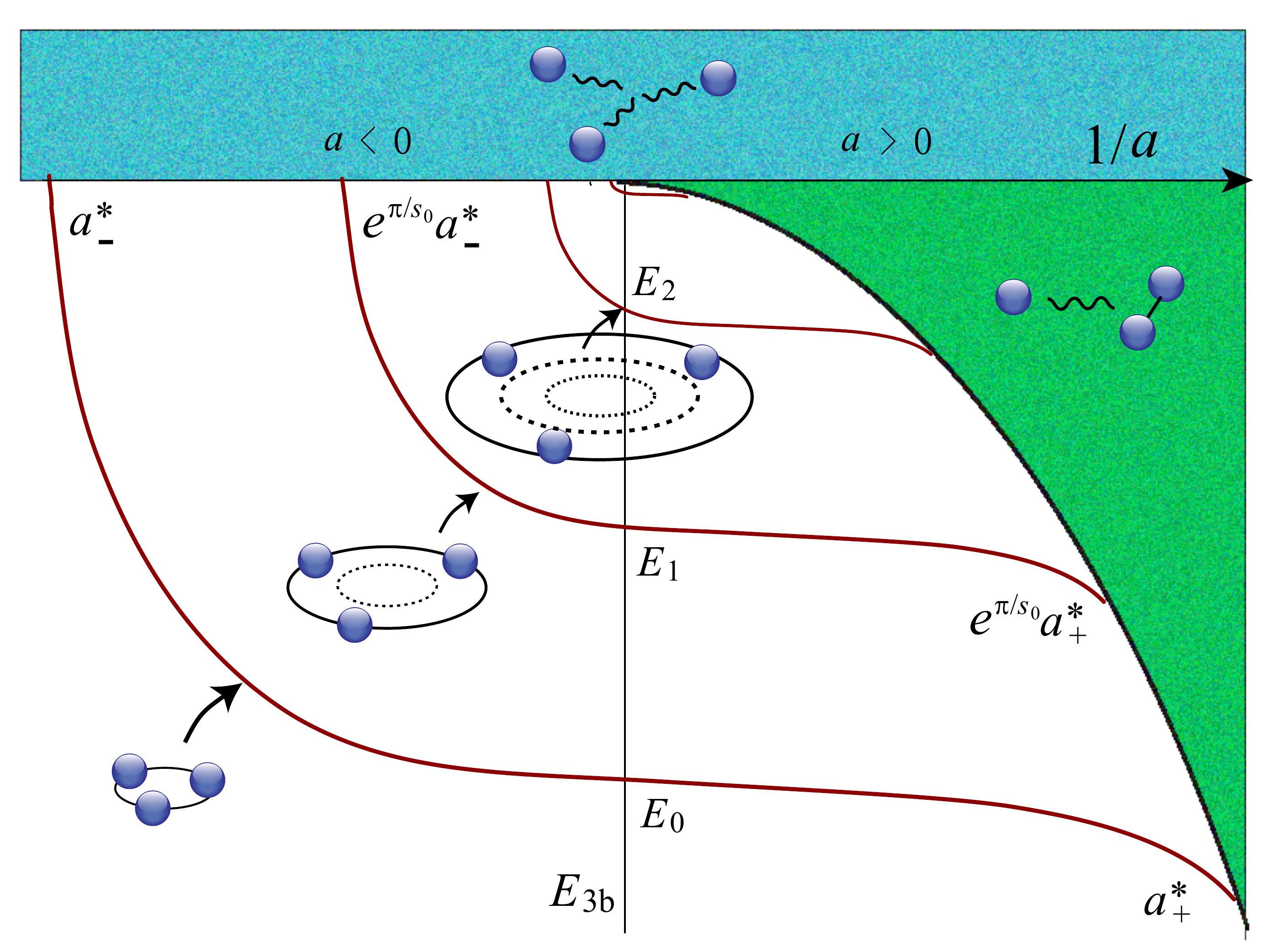}}
\caption{The Efimov spectrum in a three-body system. On the negative side of $a$, the Efimov trimer states are born 
from the three-body continuum (top blue region) as |a| increases. The first state is born at $a=a_-^*$ and consecutive ones are born 
at $a=e^{n\pi/s_0}a_-^*$. On the positive side of a where a weakly-bound two-body state exists, the 
Efimov states merge into the atom-dimer continuum each time when $a$ changes by a factor of $e^{\pi/s_0}$. The lowest merging position is 
denoted as $a_+^*$. 
\label{Fig_Efimov_Spec}
}
\end{figure}

In case of finite $|a|$, the $-1/R^2$ potential still exists but only extends to a distance $R\sim|a|$, so there will be only a finite number of Efimov states. 
As demonstrated in Fig.~\ref{Fig_Efimov_Spec}, however, when $|a|$ increases more Efimov states become bound either below the three-body breakup threshold  
or the atom-dimer threshold, for $a<0$ or $a>0$, respectively. The values of $a$ where the first Efimov state becomes bound, $a_-^*$ ($a<0$) or $a_+^*$ ($a>0$), 
as well as the position of the first three-body recombination minimum $a_0^*$~\cite{braaten2006prep,wang2013amop} at $a>0$, 
can also represent the position of the ground Efimov state and are therefore often referred to as the three-body parameter as well. 

Generally speaking, the above mentioned representations of the three-body parameter are related by the universal relations. 
For identical bosons these are known as~\cite{braaten2006prep,gogolin2008prl}
\begin{equation}
a_-^*\approx -1.501/\kappa^*,\, a_+^*\approx 0.0708/\kappa^*,\, {\rm and}\, a_0^*=0.316/\kappa^*. 
\end{equation}
In realistics systems 
such as ultracold atoms, the ground Efimov state is often ``contaminated'' by the details of short-range interactions that deviate from the $-1/R^2$ effective 
potential, so large deviations from the universal relations are often reported in the experimental observations at low scattering lengths~\cite{kraemer2006nature,knoop2009natphys,pollack2009science,ferlaino2011fewbodysys,dyke2013pra}. 

With the concept that the $-1/R^2$ Efimov potential extends to small $R$ and keeps attractive even when it is modified by short-range interactions, it is natural 
to expect that the ground Efimov state energy depends strongly on how this potential behaves from $R=0$ all the way up to $R\gtrsim r_0$, or in the very least, 
on the semiclassical WKB phase $\Phi$ in this range of the potential. It can be shown~\cite{wang2010thesis} that when $\Phi$ changes by 
$\pm\pi$ the whole Efimov spectrum is overall shifted by $n\rightarrow n\mp 1$, whereas the low-energy three-body observables are kept unchanged. 
In atomic systems as short-range interactions are strong enough to hold hundreds to thousands of ro-vibrational states already in the two-body level, 
a less than 1 percent change 
in the interactions would be sufficient to change the short-range phase by many multiples of $\pi$, which makes the prediction of the three-body 
parameter practically almost impossible. In ultracold experiments, when the magnetic field is scanned through different Feshbach resonances, 
the number of two-body bound states changes so that the WKB phase in the two-body interactions (at zero energy) changes through multiple of $\pi$'s. 
An obvious guess on the change of three-body phase is in a range greater than the change in the two-body phase. Similarly, if experiments are done 
in different atomic hyperfine states it had been expected that the change in the interactions is also significant enough to give completely 
different three-body phases. 

Another layer of complication is that in realistic three-body systems there are usually many atom-dimer breakup thresholds. 
In the hyperspherical representation 
they are manifested by atom-dimer channels, whose potential energies go asymptotically to the dimer energies. These potentials should in principle 
couple to the Efimov potential when $R\lesssim r_0$, which seems to make the prediction of a three-body parameter even less likely. 

\begin{figure}
\centerline{\includegraphics[scale=0.6]{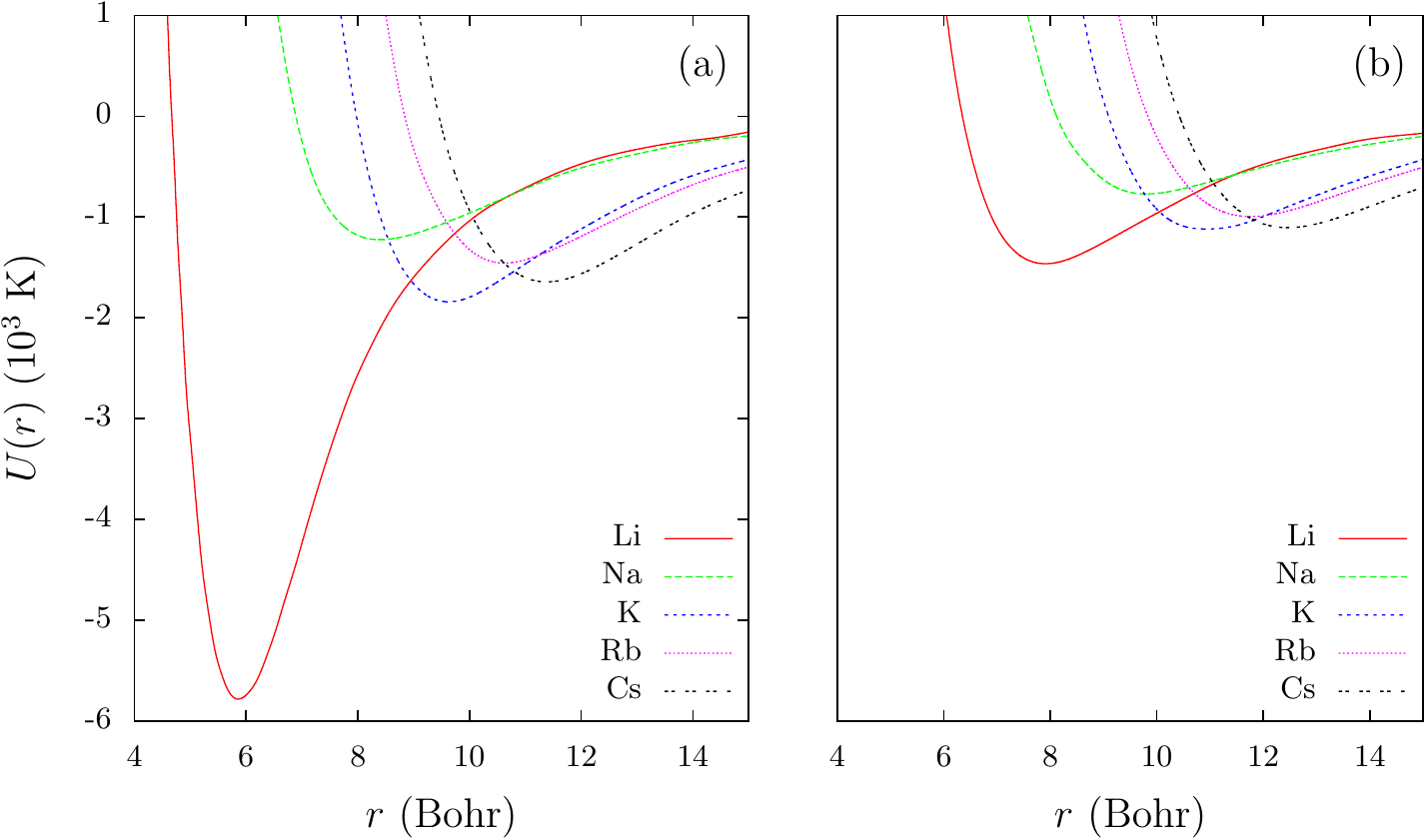}}
\caption{(a) Three-body full potential surfaces for alkali atoms that include three-body non-additive potentials~\cite{soldan2003pra}. 
(b) Three-body potentials with only the pair-wise sum of two-body potentials. The units of potentials are converted into Kelvin by $U/k_B$, where $k_B$ 
is the Boltzmann constant. In both cases the atoms are spin-polarized (quartet state) and are in the the $D_{3h}$ (equallateral triangle) geometry. }
\label{Fig_3BF}
\end{figure}

Finally as shown in Fig.~\ref{Fig_3BF}, the non-additive three-body forces for alkali atoms, which differ drastically for different atomic 
species or different atomic spin states, are often stronger than the sum of two-body interactions near the distance of chemical bond ($\lesssim 30$ Bohr). 
If three-body 
forces do contribute to the three-body short-range phase, there would be even less hope for the prediction of a three-body parameter due to the finite precision 
at which the potential surfaces can be calculated. 

Based on the above considerations, it has been generally believed~\cite{braaten2006prep,dincao2009jpb} that the three-body parameter should not carry any universal 
properties. However, the experiments with $^7$Li atoms~\cite{gross2009prl,gross2010prl} showed the first contradiction to this expectation: 
the three-body parameter was measured 
to be independent of the atoms' hyperfine state. More surprisingly, the Innsbruck group later measured $a_-^*$ for Cs near Feshbach resonances 
with quite different characters but found a persistent value of~\cite{berninger2011prl} 
\begin{equation}
\label{vdw3bp}
a_-^*\approx -10 r_{\rm vdW}, 
\end{equation} 
Moreover, the magnitude of the three-body recombination induced loss rate $L_3$ has been observed to be similar near the Efimov resonances 
in different regions of magnetic fields~\cite{ferlaino2011fewbodysys}. 
After the Innsbruck experiment, more experimental results including the earlier ones (shown in Table~\ref{Tab_3BP_Exp}), are found to be consistent 
with this ``universal'' value of $a_-^*$ when it is cast in the unit of $r_{\rm vdW}$. 
\begin{table}
\tbl{The positions of the first Efimov resonances observed in experiments.}
{\begin{tabular}{cccccc} \toprule
 &$r_{\rm vdW}$ (Bohr) & Hyperfine state & $B_0$ (G)  &$s_{\rm res}$ &$a_-^*/r_{\rm vdW}$ \\
 $^7$Li & 32.5 & $f=1$, $m_f=0$ & 894.63(24) &  0.493\footnotemark[0] & -8.12(34)\footnotemark[1]\\
   &  & $f=1$, $m_f=1$ & 738.3(3) & 0.54 & -9\footnotemark[2] \\ 
   &  &  & 736.8(2) &  & -9.17(31)\footnotemark[2] \\
\hline
$^{39}$K & 64.5 & $f=1$, $m_f=0$ & 58.92(3) & 0.11 & -14.7(23)\footnotemark[4]  \\
  &   &   &  65.67(5) & 0.11 & -14.7(23)  \\ 
  &   &   &  471.0(4) & 2.8 & -9.92(155) \\ 
  &   & $f=1$, $m_f=-1$ & 33.64(15) & 2.6 & -12.9(22)  \\
  &   &   &  162.35(18) & 1.1 & -11.3(19) \\ 
  &   &   &  560.72(20) & 2.5 & -9.92(140) \\ 
  &   & $f=1$, $m_f=1$ & 402.6 (2) & 2.8 & -10.7 (6)  \\
\hline
$^{85}$Rb & 82.1 & $f=2$, $m_f=-2$ & 155.04 & 28 & -9.24(7)\footnotemark[5]  \\
\hline
Cs & 101 & $f=3$, $m_f=3$ & 7.56(17) & 560 & -8.63(22)\footnotemark[6]  \\
  &   &   &  553.30(4) & 0.9 & -10.19(57) \\ 
  &   &   &  554.71(6) & 170 & -9.48(79) \\ 
  &   &   &  818.89(7) & 12 & -13.86(149) \\ 
  &   &   &  853.07(56) & 1470 & -9.46(28) \\ \botrule
\end{tabular}}
\begin{tabnote}
The positions of the Feshbach resonances ($B_0$) are quoted from 
the corresponding experiments. 
\end{tabnote}
\label{Tab_3BP_Exp}
\end{table}
\footnotetext[0]{\cite{jachymski2013pra}}
\footnotetext[1]{\cite{gross2009prl}}
\footnotetext[2]{\cite{pollack2009science}}
\footnotetext[3]{\cite{gross2010prl}}
\footnotetext[4]{\cite{roy2013prl}}
\footnotetext[5]{\cite{wild2012prl}}
\footnotetext[6]{\cite{berninger2011prl,ferlaino2011fewbodysys}}

The clear disagreement between the experiments and the theoretical expectation triggered a good deal of interest in understanding the physics behind 
this universality. A key question that needs to be addressed here is the insensitivity of the three-body parameter to the complexity of the 
three-body dynamics --- vibrational and rotational excitations --- at small distances, which occurs in the case of alkali atoms 
where the interactions are strong enough to have hundreds or thousands of ro-vibrational states already on the two-body level. 

This rather mysterious universality was soon understood by an effective three-body barrier in the hyperspherical coordinates whose repulsive wall 
is located near $R=2r_{\rm vdW}$. This was demonstrated 
by~\cite{wang2012prl} 
where all the two-body ro-vibrational states from rather deep van der Waals 
potentials are included in their calculations. 
Naidon, {\it et al.}~\cite{naidon2014prl} also show such a ``universal'' barrier by including only the vibrational states. Intuitively, this barrier gives 
a short-range cutoff to the Efimov potential and ``protects'' the Efimov states from the influence of three-body forces at small distances. 
This barrer leads to the suppression of the three-body amplitude inside the barrier, which is in fact not a necessary condition for 
a universal three-body parameter --- as we will discuss below for the heteronuclear systems with extreme mass ratios. The crux for 
the universality lies in the universal position of the hyperradial node or quasi-node (a node-like structure where the amplitude is small but finite) 
that can be either understood by the rise of the effective barrier or by the sharp increase in the interaction strength~\cite{wang2012prl,chin2011arXiv}. 
This nodal structure has also been interpreted as a consequence of universal three-body correlation for van der Waals interactions, and 
can be generalized to other types of short-range interactions except for square well potentials~\cite{naidon2014prl}. 

Similar to the homonuclear three-body systems, a heteronuclear three-body system can also have the Efimov effect, particularly when two of the particles are 
heavy (identical) and the scattering length between heavy ($H$) and light ($L$) particles $a_{HL}$ is large~\cite{efimov1973npa}. 
In such a system the Efimov scaling factor $e^{2\pi/s_0}$ 
is small~\cite{efimov1973npa,dincao2006pra}, excited Efimov states are therefore easy to form and this is often called an ``Efimov-favored'' scenario. 

Since the scaling behavior of the aboved-mentioned hyperradial potentials 
depends on the mass ratio~\cite{efimov1973npa,dincao2006pra}, the universality in the three-body parameter found in homonuclear systems 
cannot be readily carried over to heteronuclear systems. 
Nevertheless, Wang, {\it et al.} have shown~\cite{wang2012prlb} that the three-body parameter in a heteronuclear atomic system is also universally determined 
by $r_{\rm vdW}$ without 
being affected by the details of short-range interactions. In the ``Efimov-favored'' systems, the universality of the three-body parameter is particularly understood in a 
simple picture when the motion of the light atom is treated by the Born-Oppenheimer approximation~\cite{wang2012prlb}. 

In this case the Efimov physics is dictated by the Born-Oppenheimer potential $U^{\rm BO}(r)$, where $r$ is the distance between the $H$ atoms. 
Here the origin of the universal three-body parameter is a combination of the universal property of the 
Efimov behavior in $U^{\rm BO}(r)$ at large $r$ and that of van der Waals behavior at small $r$. 
This can be easily seen by writing $U^{\rm BO}(r)$ as 
\begin{equation}
U_\nu^{\rm BO}(r)=V_{HH}(r)+V_\nu^{\rm BO}(r),
\end{equation}
where $V^{HH}(r)$ 
is the ``bare'' $H$-$H$ interaction and $V_\nu^{\rm BO}(r)$ is the interaction induced by $L$. The index $\nu$ labels the 
quantum states of $L$ when $H$ atoms are fixed in space. The channel $\nu$ relevant to the Efimov effect is the highest state with $\sigma_g$ symmetry 
(zero angular momentum projection along the $H$-$H$ axis and the $L$ wave function is symmetric upon inversion of its coordinates). 
It is well-known that in this channel, $V_\nu^{\rm BO}(r)$ 
has the behavior 
\begin{equation}
V_\nu^{\rm BO}(r)\approx -\frac{\chi_0^2}{2m_L r^2}\hbar^2 
\label{Eq_HHL}
\end{equation}
in the range $r_{{\rm vdW}, HL}\ll r\ll |a_{HL}|$, 
where $m_L$ is the mass of $L$, $r_{{\rm vdW}, HL}$ is the van der Waals length between $H$ and $L$, 
$\chi_0\approx 0.567143$ is a universal constant. 
At $r<r_{{\rm vdW}, HL}$, $V_\nu^{\rm BO}(r)$ behaves in a complicated way due to avoided crossings with other channels, but has a magnitude on the order of the 
van der Waals energy $E_{{\rm vdW}, HL}=\hbar^2/(2\mu_{HL} r_{{\rm vdW}, HL}^2)$, where $\mu_{HL}$ is the reduced mass of $H$ and $L$. 

\begin{figure}
\centerline{\includegraphics[scale=0.7]{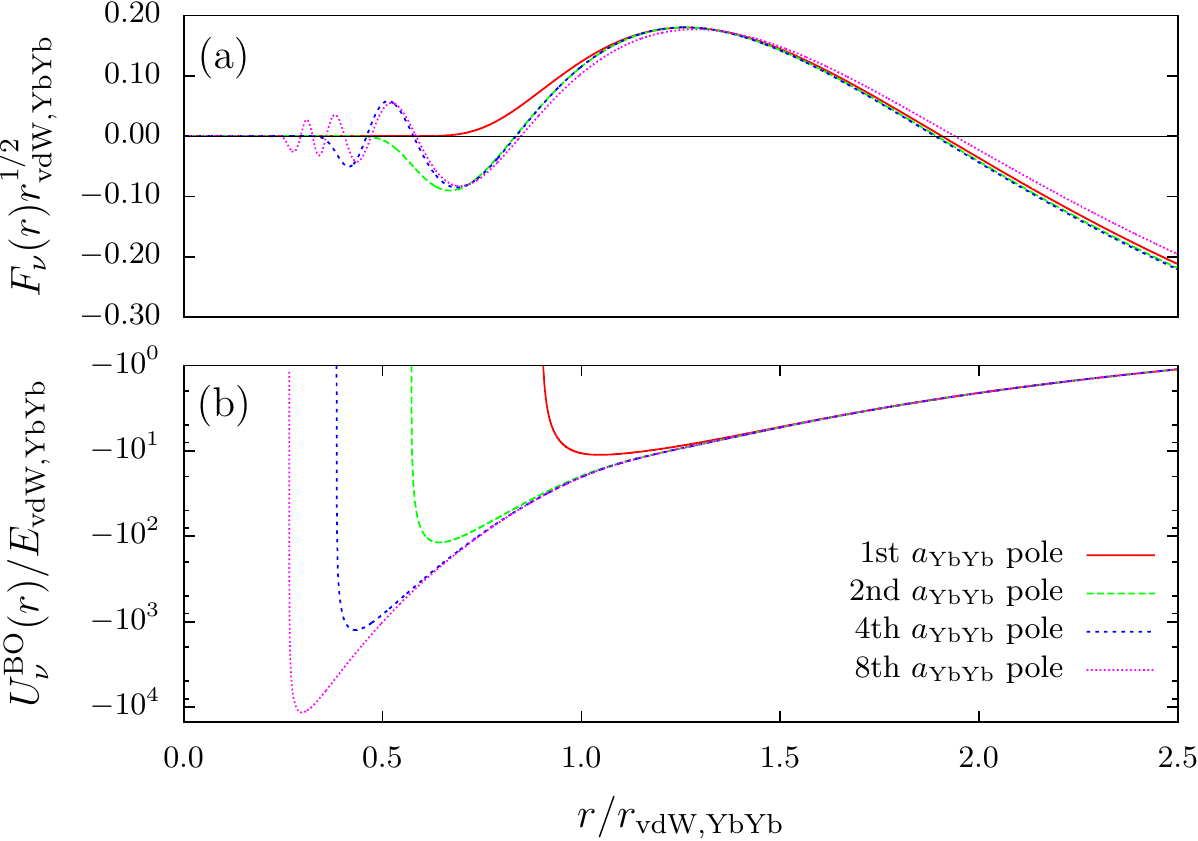}}
\caption{(a) The Born-Oppenheimer radial wave function $F_\nu(r)$ of the first excited Efimov states for YbYbLi. The shot-range cutoff of the Yb-Yb 
Lennard-Jones potential~\cite{jones1924procroysoclond} 
is tuned at the first a few scattering length poles. (b) The Born-Oppenheimer potential $U_\nu^{\rm BO}(r)$ in the Efimov channel. The short-range parameter 
correspond to those used in (a).}
\label{Fig_HHL_BO}
\end{figure}

On the other hand, $V_{HH}(r)\approx -C_{6,HH}/r^6$ in the range $r\gg r_{0,HH}$, where $r_{0,HH}$ 
is the distance where the electronic exchange interaction starts becoming significant, with $r_{0,HH}<r_{{\rm vdW}, HL}<r_{{\rm vdW}, HH}$ 
often satisfied in atomic systems. In such a scenario, the potential $U_\nu^{\rm BO}(r)$ is dominated by the $-1/r^2$ behavior when $r\gg r_{{\rm vdW}, HH}$ 
but is otherwise dominated by the van der Waals interaction $-C_{6,HH}/r^6$ when $ r_{0,HH} \ll r\ll r_{{\rm vdW}, HH}$. 
Importantly, thanks to the universal properties of 
van der Waals interaction (see Sec.~\ref{vdW_2atoms}) 
the nodal structure of the Born-Oppenheimer radial wave function $F_\nu(r)$ in the van der Waals dominant region 
is completely determined by $C_{6,HH}$, or $r_{{\rm vdW}, HH}$, without refering to the interactions 
at distances near or smaller than $r_{0,HH}$. Moreover, 
this nodal structure ``anchors'' the position of the next node in the Efimov region, which determines $E_0$, or the three-body parameter. Such ``universal'' 
nodal determination is demonstrated in Fig.~\ref{Fig_HHL_BO} for a three-atom system with extreme mass ratio --- YbYbLi. In Fig.~\ref{Fig_HHL_BO}, 
the formation of a regular nodal structure in the region $r\lesssim r_{{\rm vdW}, HH}$ is clearly shown when the van der Waals interaction 
$-C_{6,HH}/r^6$ extends to smaller distances but with $a_{HL}$ fixed. In fact, this nodal structure closely follows that in the zero-energy wave function in 
a pure $-C_{6,HH}/r^6$ potential with the same $a_{HL}$. 

Briefly speaking, the universality of the three-body parameter in an ``Efimov-favored'' system can be understood as the following: 
all the non-universal ingredients in $V_\nu^{\rm BO}(r)$ near $r\lesssim r_{{\rm vdW}, HL}$ are 
``dissolved'' by the strong van der Waals interaction 
in $V_{HH}(r)$, whereas the non-universal ingredients in $V_{HH}(r)$ are completely ``absorbed'' in $a_{HH}$. 
Using the above picture, a universal three-body parameter can be obtained analytically, which depends only on $a_{HH}$, $r_{{\rm vdW}, HH}$, 
and the mass ratio~\cite{wang2012prlb}. 

In the ``Efimov-unfavored'' heteronuclear systems where $m_H/m_L\lesssim 1$, Ref.~\cite{wang2012prlb} shows that the three-body parameter is still universal, but with 
a mechanism similar to the identical boson case, i.e., with the existence of a universal repulsive barrier. 

Finally, it should be noted that the above theoretical analyses are based on single-channel atomic interactions and are therefore not expected to 
be applicable to narrow Feshbach resonances. Experimental observations on identical bosons, 
however, have shown consistency with the universal three-body parameter for relatively narrow ones in $^7$Li 
and $^{39}$K (see Table~\ref{Tab_3BP_Exp}). 
These interesting observations are yet to be understood by future studies. 
In addition, in recent ultracold experiments on the CsCs$^6$Li system~\cite{tung2014arXiv,pires2014prl} where the Cs-Li Feshbach resonance is relatively narrow, 
the observed three-body parameter is also consistent with the theory for broad Feshbach resonance --- a prediction based on Ref.~\cite{wang2012prlb}. 
The physics behind these unexpected agreements 
is still under investigation, and may be understood by the three-body spinor models introduced in Sec.~\ref{ThreeSpinors}, 
where the multichannel character of the Feshbach resonances is properly represented.

\subsection{Three-body physics with spinors --- multichannel models}
\label{ThreeSpinors}

Although the situation of large scattering length is almost always produced by multichannel phenomena in nature, the majority of studies on Efimov physics so far are 
performed by single-channel interactions. Such treatment is adequate near broad, or open-channel dominant Feshbach resonances where studies~\cite{kohler2006rmp} 
have shown that the two-body physics in those cases can be well represented by models with single-channel interactions. 

Generally, an {\it isolated} Feshbach resonance can be characterized by two parameters: the background scattering length $a_{\rm bg}$ and the $s_{\rm res}$ parameter. 
A broad or narrow Feshbach resonance can be distinguished by $s_{\rm res}\gg 1$ or $s_{\rm res}\ll 1$. Today the question on Efimov 
physics and three-body scaling properties is more focused on the cases of non-broad Feshbach resonances where $s_{\rm res}\gg 1$ is not satisfied 
so that a single-channel treatment is expected to fail. As such Feshbach resonances are commonly accessed in ultracold experiments, it is particularly 
desirable to understand the three-body physics in such cases. 

The success of the zero-range interaction model in predicting the Efimov effect~\cite{efimov1970plb} make it attractive to extend the model to a multichannel 
version. Efimov made an early effort in this direction to understand how the Efimov effect can manifest in the presence of the nuclear spin~\cite{bulgac1976snjp}. 
Some other recent efforts focused more on reproducing the energy dependence of two-body scattering near narrow Feshbach 
resonances by 
generalizing the zero-range boundary conditions or by introducing a two-body ``molecular state'' 
into the three-body Hamilton~\cite{petrov2004prl,lee2007pra,gogolin2008prl,massignan2008pra,jona-lasinio2010prl,pricoupenko2010pra}. 
These theories 
all introduce a parametrization by the effective range $r_e$ or equivalently the $R^*$ parameter~\cite{petrov2004prl,gogolin2008prl}. For atoms with van der Waals 
interactions 
$R^*$ is defined as
$R^*=r_{\rm vdW}/s_{\rm res}$; near narrow Feshbach resonances there is further the relation $r_e\approx-2 R^*$. 
Such parametrization, though simple and elegant and is expected to work in the extreme limit $|a|\gg r_0$ and $s_{\rm res}\ll 1$, is not sufficiently capable of representing van der Waals 
physics for three atoms. 

More sophisticated multichannel zero-range models treat the closed-channel Hilbert space in equal footing with 
the open-channel space, 
rather than simply a bound ``molecular state''. A relatively straightforward way to do this is to write the three-body wave function $\Psi$ as 
a superposition of open- and closed-channel components $\psi_{\rm open}$ and $\psi_{\rm close}$, and match $\Psi$ separately in 
the open- and closed-channel zero-range boundary conditions~\cite{sorensen2012pra,sorensen2013jpb,sorensen2013fewbodysys}. 
A more rigorous way is to build the internal degrees of freedom --- 
such as spins --- directly to individual atoms, as the multichannel zero-range model by~\cite{mehta2008pra}. 

Thanks to their simplicity, the multichannel zero-range models have recently been applied to fit three-body recombination losses in ultracold 
experiments~\cite{sorensen2013fewbodysys} and to study spinor condensates~\cite{colussi2014fewbodysys}. It should be noted, however, that the zero-range boundary condition 
for the closed-channel wave function implies a weakly-bound two-body Feshbach state in the closed channel, which is rarely the case for realistic 
systems. In addition, the multichannel zero-range models still have the Thomas collapse~\cite{sorensen2013fewbodysys}, which also limits their predictive power 
for ultracold experiments. 

To better understand the role of van der Waals physics in ultracold collisions of three atoms, it is desirable to have a multichannel model that has 
finite-range van der Waals interactions 
built in. Moreover, the connection found by~\cite{wangj2012pra} 
between a $d$-wave two-body van der Waals state and a three-body state 
suggests that a correct characterization of the high-lying two-body van der Waals spectrum is also important for predicting ultracold three-body physics in atoms. 

Recently, significant progress has been made by~\cite{wang2014natphys,wangUnpub} 
down this road, 
where the authors have developed a three-body multichannel model with van der Waals interactions. By allowing atoms in their model to carry multiple spin states and 
interact via spin-coupled van der Waals potentials, the authors have successfully reproduced/predicted many experimentally observed three-body 
loss features without fitting parameters, particularly in the case of Cs where three-body physics has been extensively 
studied in Innsbruck experiments~\cite{ferlaino2011fewbodysys}. 
After the groundbreaking discovery of a universal three-body parameter, this theoretical work shows that without addressing the unknown 
short-range details, ultracold three-body physics in atoms can be quantitatively predicted. 
This pushes the front of few-body research to a much further position. 

In Wang, {\it et al.}'s work, three atoms are represented by the Hamiltonian~\cite{wang2014natphys}  
\begin{align}
H=T-(\mu_{s_i}+\mu_{s_j}+\mu_{s_k})B+u_{s_i}+u_{s_j}+u_{s_k}\\\nonumber+\sum_{s_i,s_j,s_k}|s_k\rangle\langle s_k| 
\otimes |s_i s_j\rangle {\cal S}^\dagger v_{s_i s_j,s_i' s_j'}{\cal S}\langle s_i' s_j'|.
\end{align}
Here $T$ is the kinetic energy for the relative motion of three atoms, $|s_{i,j,k}\rangle$ is the spin state of atom $i$, $j$, or $k$, 
which can take $|a \rangle$, $|b \rangle$, $|c\rangle$, ... in a multi-spin model.  
Also, $\mu_{s_i}$ is the magnetic moment for an individual atom, and $u_{s_{i,j,k}}$ is the zero-field, single-atom energies that represent the hyperfine splittings. 
The pairwise two-body interactions $v_{s_i s_j,s_i' s_j'}$ are in a symmetrized two-body spin basis ${\cal S} |s_i s_j\rangle$ (${\cal S}$ is the symmetrization operator), 
with 6-12 Lennard-Jones potentials 
in the diagonals. Thanks to the universal properties of two-body van der Waals physics~\cite{gao2000pra}, the near-threshold two-body spectrum can be reasonably 
reproduced by the Lennard-Jones potentials that are deep enough for a few bound states. By including more spin states 
this Hamiltonian is capable of representing complicated Feshbach physics in cold atoms. For instance, isolated Feshbach resonances can be 
represented by including two spin states for each atom, whereas overlapping Feshbach resonances can be represented by three spin states. 

With this model, Wang, {\it et al.} have studied an interesting case in Cs where an Efimov doublet was observed near a narrow 
$g$-wave Feshbach resonance~\cite{berninger2011prl,ferlaino2011fewbodysys}. 
Their calculations successfully reproduced the positions of the Efimov resonances in the doublet and showed how the magnitude of the loss rates 
can be predicted correctly when a 2-spin model is replaced by a 3-spin model, where the overlapping character of the $g$-wave Feshbach resonance 
is better represented~\cite{wang2014natphys}. 
Moreover, even though the theory agrees with the experiment in that the positions of the Efimov resonances in this case are all near 
the universal value, Wang, {\it et al.}'s theory further points out that the universal resonance positions are actually a consequence of $a_{\rm bg}$ being near the universal 
value ($-10 r_{\rm vdW}$) and should not be considered as general~\cite{wangUnpub}. 

The above mentioned multichannel three-body models can also be applied to heteronuclear atomic systems. For instance, to study the CsCsLi Efimov physics in the ultracold 
Cs-Li admixtures realized in the Chicago group~\cite{tung2014arXiv} and the Heidelburg group~\cite{pires2014prl}, a 2-spin model can be used to 
treat the CsLi interactions~\cite{tung2014arXiv}. With the great variety of heteronuclear combinations that are accessible to future experiments, theoretical studies with 
multichannel models are just in the beginning. In this context, many questions on the van der Waals universality still remain to be answered. 

Finally, it should mentioned that the magnetic Feshbach resonances are by far not the only place where multichannel physics is important. 
In the emerging fields of ultracold studies where light is 
used to manipulate the atomic interactions, such as the optical Feshbach resonance~\cite{chin2010rmp} 
and synthesized spin-orbit interactions~\cite{zhai2012intjmpb}, new, interesting few-body phenomena due to multichannel physics is already start being 
predicted or observed (see Sec.~\ref{LongRange}). 

\section{van der Waals physics in the chemical reaction of compact molecules}\label{vdW_2mol}\index{index!multiple indexes} 

We have been discussing the weakly bound states of three atoms, or the collisions of an atom with a weakly bound dimer. Now we turn to collisions or chemical reactions of ordinary molecules, for example, in their ground vibrational state.  Such molecules are much more complex than atoms, having additional degrees of freedom, and their chemical reaction dynamics provides an excellent example of few-body dynamics.  While we may expect that molecular dynamics is too complex to show any universality, we will show here that there are some cases, perhaps even widespread, where universality is found, that is, insensitivity of the reaction rate coefficient to the details of short-range chemical interactions.

It is well-known that some chemical reactions are described by a universal classical model when there is a unit probability of a reaction occurring if the two reactant species come in close contact with one another.   The collision cross section is then determined by the long range potential: it is only necessary to count those classical trajectories of the colliding species that are captured by the long range potential such that the collision partners spiral in to close contact with each other.  
The Langevin model~\cite{langevin1905acp} is a good example of this for ion-molecule reactions, and the related Gorin model~\cite{gorin1938actphys,gorin1939jcp} describes barrierless reactions between neutral chemical species that collide via a van der Waals potential. \cite{fernandez-ramos2006cr} describes the Langevin and Gorin models in the context of chemical kinetics. 

In the low temperature regime such classical capture models need to be adapted to the quantum mechanical near-threshold dynamics governed by quantum threshold laws and resonant scattering.  This can be readily done based on QDT treatments of neutral~\cite{gao1996pra} or ionic reactions~\cite{jachymski2013prl}. 
We concentrate on the neutral case here, which has also been reviewed by~\cite{quemener2012cr}.  
It is important to note that QDT theories are not unique, but can be implemented in different ways, in either their analytic or numerical versions, depending on how the reference solutions to the problem are set up.  We concentrate here on the particular QDT implementation of Mies~\cite{mies1984jcpa,mies1984jcpb,julienne1989josab}.  Gao~\cite{gao2008pra,gao2010prla,gao2010prlb,gao2013pra} 
obtains similar results with his implementation of the theory.

As discussed in Sec.~\ref{MQDT}, the 
essence of QDT for molecular reactions is the separation of the interactions into short range and long range regions, with the former characterized by a few quantum defect parameters that allow for chemical reaction at short range.  Generalizing the classical Gorin or Langevin models to the quantum threshold is especially simple in a QDT framework~\cite{idziaszek2010pra}.  The approach of the two colliding species is governed by long range quantum dynamics, whereas the short range region is characterized by perfectly absorbing, or ``black hole'' boundary conditions that allow no backscattering of the incoming partial wave from the short range region.  On the other hand, the long range region can quantum reflect the incoming flux in a way strongly dependent on partial wave $\ell$, allowing for quantum tunneling to short range in the case of centrifugal barriers when $\ell \ge 1$.  Following Gao~\cite{gao2010prla} we will call QDT 
models with such ``black hole'' boundary conditions quantum Langevin (QL) models.

In general, the short range region may not exhibit unit reactivity, but may reflect some of the incoming flux back into the long range entrance channel.  The interference of incoming and outgoing flux establishes a phase shift of the long range standing wave that is related to the scattering length.  In this ``grey hole'' partially reactive case, two dimensionless quantum defect parameters are needed: 
one, $s$, related to the phase shift and one, $0 \le y \le 1$, 
related to the short range reactivity.  The probability $P$ of short range reaction is $P=4y/(1+y)^2$~\cite{jachymski2013prl}.  The rate constants for elastic scattering or for inelastic or reactive scattering loss from the entrance channel are determined from the diagonal $S$-matrix element $S_\ell(E)=e^{2i\eta_\ell(E)}$, which can be specified for any partial wave and $E$ by the complex phase shift $\eta_\ell(E)$, or alternatively, by a complex energy-dependent scattering length $\tilde{a}_\ell(E)$ defined by~\cite{idziaszek2010prl}:
\begin{equation}
  \tilde{a}_\ell(E) = -\frac{\tan{\eta_\ell(E)}}{k} = \frac{1}{ik} \frac{1-S_\ell(E)}{1+S_\ell(E)}
  \label{Eq:complex_a}
\end{equation}
~\cite{idziaszek2010prl} 
give the universal van der Waals $s$- and $p$-wave $\tilde{a}_\ell(E)$ in QDT form, which reduce in the $\kappa \to 0$ limit to (in van der Waals $\bar{a}$ units): 
\begin{equation}
  (\mathrm{a}) \,\,\,\tilde{\alpha}_{\ell=0} \to s +y \frac{1+(1-s)^2}{i+y(1-s)} \,\,\,\,\,\,\,(\mathrm{b}) \,\,\, \tilde{\alpha}_{\ell=1} = -2\kappa^2 \frac{y+i(s-1)}{ys+i(s-2)} \,.
\end{equation}

In the specific ``black hole'' case of the QL model, where $P=y=1$, the complex scattering lengths take on an especially interesting universal threshold form, independent of $s$ and with equal magnitudes for their real and imaginary parts: $\tilde{\alpha}_{\ell=0} = 1-i$ and $\tilde{\alpha}_{\ell=1} = -\kappa^2 (1+i)$.  The corresponding elastic and reactive/inelastic loss rate constants are determined respectively from $|1-S_\ell(E)|^2$ and $1-|S_\ell(E)|^2$.  The loss rate constants for the lowest contributing partial waves near the $k \to 0$ threshold are
\begin{equation}
   (\mathrm{a}) \,\,\,K_{\ell=0}^\mathrm{loss} = 4\pi g \frac{\hbar}{\mu} \bar{a} \,\,\,\,\,\,\,\, (\mathrm{b}) \,\,\,K_{\ell=1}^\mathrm{loss} = 40.122  g \frac{\hbar}{\mu} k^2\bar{a}^3 \,,
   \label{Eq:vdWUrate}
\end{equation}
where the symmetry factor $g=2$ in the case of identical particles in identical internal states and $g=1$ otherwise; identical bosons only have even $\ell$ collisions and identical fermions only odd ones, whereas nonidentical particles have both.  The numerical factor in the $p$-wave expression is $\Gamma(\frac14)^6/(12\pi \Gamma(\frac{3}{4})^2) \approx 40.122$, where all three components of the $p$ wave are summed, assuming them to have identical complex scattering lengths, as would be the case for a rotationless molecule (total angular momentum $J=0$).  Thus, the threshold thermally averaged universal van der Waals 
rate constants for $s$-wave collisions is the constant expression in Eq.~\ref{Eq:vdWUrate}(a), and for $p$ wave collisions of identical fermions it is 
\begin{equation}
   K_{\ell=1}^\mathrm{loss} \approx 1512.6 \frac{k_\mathrm{B}T}{h} \bar{a}^3 \,,
    \label{Eq:vdWUprate}
\end{equation}
where $k_\mathrm{B}$ is the Boltzmann constant and $T$ is the temperature.  Thus, threshold $s$-wave collisions for ``black hole'' collisions are universally specified by the vdW length $\bar{a}$ and $p$-wave ones by the van der Waals volume $\bar{a}^3$. 

The predictions of these universal rate constants for the QL model agree well with measurements on threshold chemical reaction rates of $^{40}$K$^{87}$Rb fermions in their $v=0$, $J=0$ state.  The $p$-wave expression in Eq.~\ref{Eq:vdWUprate} agrees within experimental uncertainty when the fermions are in a single spin state, verifying the linear variation of rate with $T$.  The $s$-wave expression in Eq.~\ref{Eq:vdWUrate}(a) agrees within a factor of 2 or better when the fermions are in different spin states or when they react with $^{40}$K atoms~\cite{idziaszek2010prl}.  

~\cite{quemener2010praa,ni2010nature} 
developed an approximate threshold model to extend the Gorin model when an electric field is 
introduced 
to cause the molecules to have a laboratory-frame dipole moment.   ~\cite{idziaszek2010pra} 
extended this method using a hybrid numerical quantum defect theory based on a sum of van der Waals 
and dipolar long range potentials to predict the variation of reaction rates of identical fermions with electric field, in excellent agreement with experiment.  These quantum defect treatments can readily be adapted to quasi-1D (``tube'' geometry with tight confinement in two directions)~\cite{olshanii1998prl} 
or quasi-2D (``pancake'' geometry with tight confinement in one direction)~\cite{petrov2001pra} 
collisions of reduced dimensionality D due to quantum confinement by optical lattice structures.  The three-dimensional (3D) boundary conditions due to short range interactions is incorporated in the QDT 
parameters $s$ and $y$, and the asymptotic boundary conditions can be taken to be appropriate to the situation of reduced dimension due to quantized tight confinement~\cite{naidon2007njp}.  ~\cite{micheli2010prl} 
extend the universal vdW rate constants in Eqs.~\ref{Eq:vdWUrate}(a) and~\ref{Eq:vdWUprate} to quasi-1D and quasi-2D geometry, and additionally develop numerical methods for ``black hole'' collisions in quasi-2D geometry when the molecules have a dipole moment; ~\cite{quemener2011pra} 
also give similar calculations for KRb, and ~\cite{quemener2011prab} and ~\cite{julienne2011pccp} 
discuss the universal collisions of polar molecules containing other alkali-metal atoms.  These calculations explain experiments with $^{40}$K$^{87}$Rb fermions in ``pancake'' quasi-2D geometry, verifying earlier predictions~\cite{buchler2007prl} that aligning the dipoles perpendicular to the plane of the ``pancake'' suppresses the reaction rate, instead of increasing rapidly with dipole strength as in ordinary three-dimensional collisions.

QDT models can be readily extended to include more partial waves as temperature $T$ increases.  Both ~\cite{gao2010prla,gao2010prlb} and ~\cite{jachymski2013prl} 
have done this to get universal QL rate constants for both the van der Waals 
and the ion-atom cases ($N=$ 6 and 4), including the quantum effect of centrifugal barrier tunneling. The universal QL model goes to the classical Gorin or Langevin models when the temperature goes above a characteristic temperature associated with the long range energy scale.  The QDT model can also be implemented for the more general ``grey hole'' case, where both $s$ and $y$ are needed.  Jachymski {\it et al}~\cite{jachymski2013prl} apply the general QDT theory to explain the merged beam experiments on the Penning ionization of $^3$S$_1$ metastable He atoms with Ar atoms~\cite{henson2012science} up to collision energies on the order of 10 K.  In this case, where both atoms are in $S$ states and there is a single potential, the contributions from all partial waves, including a prominent shape resonance for $\ell=5$, can be explained by a single $s$-wave complex scattering length parameterized by $s=3$, $y=0.007$.  In the more general case of molecules with anisotropic potentials, we can expect to need $\ell$-dependent QDT parameters.   Gao's discussion of the general case~\cite{gao2008pra,gao2010prla} also describes the role or resonance states, which allow incoming and outgoing scattering flux to make multiple passes between the inner and outer regions, quite unlike the ``black hole'' case, where resonances are suppressed by the total absorption at short range.

The review by ~\cite{quemener2012cr} 
examines the applicability of the universal QL model to a variety of molecular collisions governed by the van der Waals 
potential.  They review work on the vibrational quenching collisions of atom $A$ with $AB$($v$) or $A_2$($v$) molecules, where $A$ and $B$ represent alkali-metal atoms and $v$ is an excited vibrational level.  
Both measured $AB$($v$) quenching in magneto-optical traps around 100 $\mu$K (see their Fig.~21) 
and calculated threshold quenching of $A_2$($v$) 
(see their Fig.~41) 
tend to be within around a factor of 2 to 4 from the predictions of universal QL model.  This suggests that vibrational quenching in general tends to have a high probability of a short range quenching event, that is, has a relatively large $y$ not far from unity.  The question that remains to be answered is why the quenching rate coefficients for different species and $v$ levels fluctuate by several factors from the predictions of the QL model.

Finally, it is important to note that real molecular collisions are likely to involve a very dense set of resonance states associated with the various vibrational and rotational degrees of freedom in the molecule.  The Bohn group~\cite{mayle2012pra,mayle2013pra,croft2014pra} has described how the density of resonances are so high that statistical random matrix theories are needed to describe their effects even at ultralow temperatures.  Recent cold atoms experiments with the strongly magnetic dipolar atomic species Er have uncovered a high density of Feshbach resonance states characterized by a Wiger-Dyson distribution of resonance spacings that is a characteristic of random matrix theory~\cite{frisch2014nature}; similar experiments are being carried out with Dy atoms~\cite{baumann2014pra}.  If there are many resonances within the thermal spread $k_\mathrm{B}T$, the molecules are predicted to 
stick together in a long-lived collision complex with a universal collision rate given by the QL model~\cite{mayle2013pra}. If non-reactive molecules in such a collision complex should undergo short range loss processes with unit probability and never returned outgoing flux to the entrance channel, then they would undergo loss processes at the same rate as a highly reactive collision. The question of the magnitude and control of collision rates for the quenching of internal vibration, rotation, or spin for ultracold molecules having dense sets of resonance levels, as well as the nature of their three-body collisions, remains an open research area experimentally and theoretically.

\section{Dipolar physics for two and three atoms}\label{Dipole}\index{index!multiple indexes} 

As discussed in the previous sections, 
the recent experimental development in producing ultracold ground-state polar molecules~\cite{quemener2012cr} or ultracold highly magnetic 
atoms~\cite{hancox2004nature,werner2005prl,griesmaier2005prl,stuhler2005prl,lahaye2007nature,kanjilal2008pra,
lahaye2008prl,koch2008naturephys,connolly2010pra,lu2011prl,lu2012prl,aikawa2012prl,aikawa2014prl,petrov2012prl,frisch2014nature} 
has stimulated a lot of interest in the study of few-body systems with the long-range dipolar interactions. To the simplest level, dipolar atoms and molecules 
can both be modeled as point dipoles with a ``permanent'' dipole moment that can be aligned by external fields. 
As has been discussed in Sec.~\ref{vdW_2mol}, however, the elastic properties of such models do not hold for reactive molecules due to the highly inelastic 
chemical processes that occur when molecules are close together, 
and the random matrix theory arguments~\cite{mayle2012pra,mayle2013pra,croft2014pra} predict that even for non-reactive molecules the elasticity is practically non-observable. 
We therefore restrict the application 
of our discussion below to atoms only where the elasticity of the short-range interactions is still preserved at current experimental conditions. 

\subsection{Universal dipolar physics for two field-aligned dipoles}
\label{2dipoles}

Generally speaking, the dynamics of dipoles include rotation of individual dipoles where the angular momentum of a dipole is exchanged with its environment.  
In practice, however, it is highly challenging to include the rotational degrees of freedom in studies of few-dipole physics even for point dipoles. 
Nevertheless, from the point of view of universal physics it is more preferential to ``freeze'' the rotation of individual dipoles by orienting them with external fields and study the scaling behavior of dipolar systems with the anisotropic, long-range interactions. 

For oriented dipoles, their interaction potential can be simply written as
\begin{equation}
\label{Eq_Dipole_Int}
V_{\rm dd}(\boldsymbol{r})=\frac{2d_l}{\mu_2}\frac{1-3({\boldsymbol{\hat{e}}\cdot \boldsymbol{r}})^2}{r^3},
\end{equation}
where the dipole length $d_l$ is the characteristic length scale for the dipolar interaction~\cite{bohn2009njp}, 
which connects to the induced dipole moment $d_m$ by 
\begin{equation}
d_l=\mu_2 d_m^2\hbar^2
\end{equation}
and is tunable by an external fields~\cite{roudnev2009jpb}. The unit vector $\boldsymbol{\hat{e}}$ is along the direction of the external field. 
For magnetic dipolar atoms $d_l$ can saturate to as large as a few hundred Bohr, whereas for alkali dipolar molecules the saturation 
value is between $10^3$ Bohr 
to $10^5$ Bohr. 
Without a short-range cutoff,  the dipole potential is too singular at the origin --- the system collapses again due to the 
``fall-to-the-center'' problem~\cite{perelomov1970theomathphys}. 
In reality the ``short-range'' interaction always deviates from Eq.~(\ref{Eq_Dipole_Int}) 
and is regular at the origin. A non-universal 
short-range cutoff funtion $f_c(r)$ needs to be introduced in theoretical studies such that 
\begin{equation}
f_c(r)V_d(\mathbf r)\rightarrow {\cal O}(r^0), r\rightarrow 0;\; f_c(r)V_d(\mathbf r)\rightarrow V_d(\mathbf r), r\gg r_0.
\end{equation}

In spite of the non-universal behavior at small distances, universal scaling properties with the strength of the dipolar interaction $d_l$ are in the central of the study. 
Ref.~\cite{wang2012pra} shows that unlike short-range interactions, the dipole potential leads to a low-energy expansion in the scattering phase shift, 
whose real part of $\delta_\ell$ characterizes the cross-section where $\ell$ is not changed during a collision and has the form 
\begin{equation}
{\rm Re}[\delta_\ell(k)]=-a_\ell k-b_\ell k^2-V_\ell k^3 + O(k^4).
\end{equation}
for any partial wave $\ell$, with the power of wavenumber $k$ increases by 1 for consecutive terms. 
Moreover, it is found that the expansion coefficients 
may or may not have short-range dependence according to $\ell$~\cite{wang2012pra}. For $\ell=0$ all the coefficients are short-range dependent. For $\ell=1$, however, 
short-range dependence only starts from the $k^3$ term and beyond. Therefore, for bosons non-universal behavior is generally expected in low-energy scattering, 
whereas for fermions universal expressions for scattering observeables can be derived~\cite{bohn2009njp,wang2012pra}. 
The imaginary part of $\delta_l$ characterizes $\ell$-changing collisions, which has $k^2$ in the leading order 
and has the leading coefficient that has been analytically derived~\cite{wang2012pra}. 

In analogy to the magnetic Feshbach resonance, formation of dipolar resonances by tuning $d_l$ is also of great interest. 
It is known that the dipolar couplings between different $\ell$ lead to both shape and Feshbach resonances~\cite{deb2001pra,ticknor2005pra,ticknor2007pra,kanjilal2008pra,roudnev2009pra,roudnev2009jpb} and 
the relative positions between these resonances are universal~\cite{kanjilal2008pra,roudnev2009pra,roudnev2009jpb} --- 
i.e., independent of short-range details. Due to the anisotropy of the dipolar interaction. However, the different partial wave characters that dipolar resonances 
bear make them line up with $d_l$ quite irregularly. 
Nevertheless, such knowledge is still important for the development of many-body dipolar theory, 
where the renormalization theory can be readily applied and system independent properties can be derived~\cite{beane2001pra}. 

Being the connector between microscopic and macroscopic phenomena, few-body studies, in particular two-body studies, play the role of developing simple models 
that can be used in studies of many-body physics. To this end, zero-range models that can correctly reproduce low-energy scattering properies have been developed 
in both 2D~\cite{kanjilal2006pra,shih2009pra,dincao2011pra} and 3D~\cite{yi2000pra,yi2001pra,derevianko2003pra,derevianko2005pra}. 
In view of stable dipolar gases against inelastic collisions, elastic two-dipole physics in 2D or quasi-2D geometries has been studied in great details in 
Refs.~\cite{ticknor2009pra,ticknor2010pra,armstrong2010epl,ticknor2011pra,dincao2011pra,volosniev2011jpb,ticknor2012pra,zinner2012fbs,koval2014pra}. 

For studies of Efimov physics, however, zero-range models do not 
have enough information in the scaling behavior of near-threshold bound states, which is crucial for studies of the scaling laws. Numerical study based on 
the potential in Eq.~(\ref{Eq_Dipole_Int}), however, indicates that the characteristic size of the near-threshold dipolar states scales like $d_l$ and their 
energies scales with $-1/d_l^2$~\cite{wang2012pra}. 

In studies on the transition from the weak dipole regime ($d_l\lesssim r_0$) to the strong dipole regime ($d_l\gg r_0$), the concept of angular momentum mixture 
is often used to indicate the level of anisotropy of a dipolar state. Although this characterization is useful in connecting the ordinary molecular physics 
to the dipolar physics when the dipolar interaction is not very strong (or the dipoles are not strongly aligned), it becomes less useful in the strong dipole 
regime where lots of angular momenta make contributions and dipolar states take characters of ``pendulum'' states~\cite{wang2012pra}. Nevertheless, 
the expectation value of the square of the angular momentum $\langle\hat{L}^2 \rangle$ in this regime follows a universal scaling 
\begin{equation}
\langle\hat{L}^2 \rangle\propto \sqrt{\frac{d_l}{r_0}}(2\nu+m_\ell+1),
\end{equation}
where $\nu$ is the vibrational quantum number of the pendulum states and $m_\ell$ is the magnetic quantum number. 

\subsection{Universal three-dipole physics}

The universal properties found in two-dipole physics have clearly stimulated interest 
in the universal physics for three dipoles. 
To begin with, we would first like to point out the non-trivial aspects in three-dipole physics, particularly those relevant to Efimov physics: 
\begin{itemize}
\item The prediction of the Efimov effect was based on isotropic interactions where the total orbital angular momentum $J$ is conserved. In the case of dipoles, however, 
it is not clear how the couplings between partial waves may impact the Efimov effect. 
\item The Efimov effect also assumes a scattering length {\it only} in the $s$-wave collisions, whereas the long-range dipolar interaction brings in equally defined 
scattering lengths 
for each partial wave.  
\end{itemize}
With the above question marks, a clear, definitive answer to the three-dipole Efimov effect can only be given by a quantitative three-body theory 
where dipolar interactions are properly treated. To this end, the hyperspherical method would be a good choice for the study, 
where the existence of the Efimov effect can be directly read off from the long-range scaling behavior of the hyperradial potentials~\cite{wang2010thesis}. 

By numerically solving the Schr{\"o}dinger equation in the hyperspherical coordinates, Wang, {\it et al.} showed~\cite{wang2011prl} 
that the Efimov effect does exist for three bosonic dipoles near an $s$-wave dipolar resonance where the $s$-wave scattering length $a_s$ goes through a pole. 
It is also found that the dipolar Efimov states follow the same scaling properties as the non-dipolar states. 
More interestingly, contrary to a common expectation that the three-body parameter is the most likely 
non-universal because of the irregular pattern in the two-dipole spectrum (see Sec.~\ref{2dipoles}), the ground Efimov state energy is found to be 
universally determined by $d_l$ without referring to the short-range details. 
This finding is similar to what has been discussed in Sec.~\ref{3BP}, 
but was discovered before the concept of universal three-body parameter being built. 
In this case, the universal three-body parameter is also manifested 
by a repulsive barrier in the hyperradial potential, but appears 
near $R=0.7d_l$. 
The dipolar Efimov state is still $s$-wave dominant, the three dipoles in this state therefore prefer to stay far away so that its size is much 
bigger than $d_l$. 
At the dipolar resonance, the position of the ground Efimov state, given by the real part of its energy, scales like 
\begin{equation}
{\rm Re}(E_0)\approx 0.03\frac{\hbar^2}{m d_l^2}. 
\end{equation}
Interestingly, applying the universal relation between $E_0$ and the Efimov resonance position $a_-^*$ leads to 
\begin{equation}
a_-^*\approx -9 d_l, 
\end{equation}
which has a 
numerical factor very close to that in the van der Waals case [see Eq.~(\ref{vdw3bp})]. In addition to the position of the ground Efimov state, its 
width also shows an $1/d_l^2$ overall scaling behavior, which leads to a quasi-universal scaling for the decay rate of an Efimov state. 
This decay is characterized by the $\eta_*$ parameter~\cite{braaten2006prep} which can be expressed as 
$\eta_*=(s_0/2)[{\rm Im}(E_0)/{\rm Re}(E_0)]$ at $a=\infty$. 
Based on the numerical results~\cite{wang2011prl} the dipolar $\eta_*$ has a value around 0.084. 
However, the width of the dipolar Efimov state is less universal than its position ---  
with a finite sample of values in $d_l/r_0\gg 1$ and some variations in the form of short-range cutoff, 
\cite{wang2011prl} shows that $\eta_*$ can vary up to 60\% depending on which dipole resonance the three-body calculation was carried out for. 

The stability of ultracold bosonic dipolar gases strongly depends on the dipolar Efimov physics discussed above. In an early study, Ticknor and Rittenhouse~\cite{ticknor2010prl} performed scaling analysis on three-dipole loss rates and derived $a_s^4$ scaling law when $|a_s|\gg d_l$ and $d_l^4$ scaling law when $|a_s|\ll d_l$. Based on the knowledge in Efimov physics and the scaling behavior of two dipoles discussed in Sec.~\ref{2dipoles}, Wang, {\it et al.}~\cite{wang2011prl} have obtained the same scaling laws. These scaling laws indicate significant losses for a 3D dipolar gas in the strong dipole regime. 
Neverthless, a stable dipolar gas can be prepared in optical lattices or reduced-dimension traps~\cite{danzl2009njp,danzl2010naturephys,chotia2012prl,hummon2013prl,yan2013nature,takekoshi2014arXiv}. 

\begin{figure}
\centerline{\includegraphics[scale=0.3]{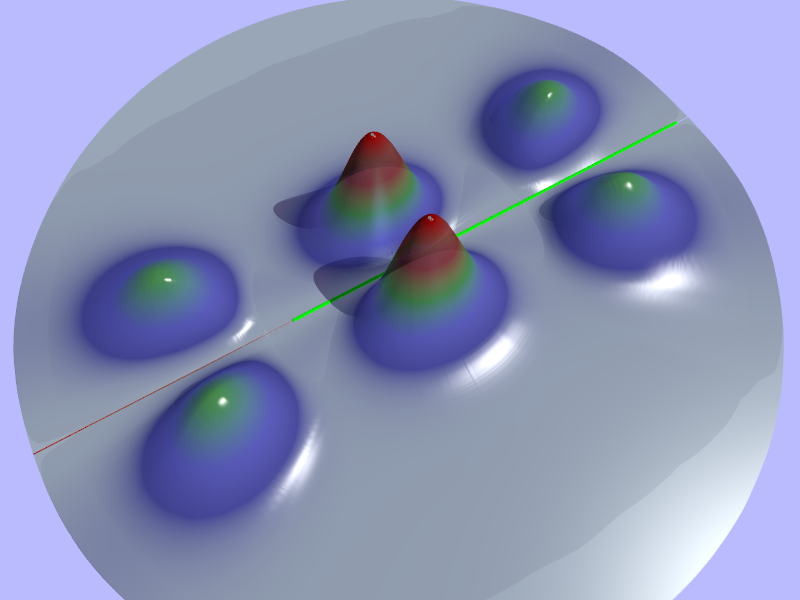}}
\caption{A cut of the geometrical distribution of the fermionic dipoles in a universal three-body dipolar state along the direction of the external aligning field, indicated by the greern line in the middle of the plane.} 
\label{Fig_Dipole}
\end{figure}

For fermionic dipoles where $s$-wave scattering is absent, Ref.~\cite{wang2011prlb} showed the existence of a $p$-wave scattering length does not modify 
the long-range scaling behavior. The threshold law for three fermionic dipoles is therefore the same as the non-dipole case~\cite{esry2001pra}. Near a $p$-wave 
dipolar resonance, however, different from non-dipole fermions where no three-body state exists, there is one, and just one universal three-dipole state with binding energy $E_{3b}$ that scales with $d_l$ as
\begin{equation}
E_{3b}\approx 160/m d_l^2.
\end{equation}
A three-body resonance is therefore expected between two two-body $p$-wave resonances. In contrast to the bosonic dipoles, the binding of fermionic dipoles 
is mainly from the anisotropy of the dipolar interaction and therefore  has a size smaller than $d_l$. This dipolar state 
has a preferential spatial distribution as shown in Fig.~\ref{Fig_Dipole}. In the most probable configuration the bond lengths 
were numerically found to be universally determined by $d_l$ as $0.14 d_l$, $0.14 d_l$, and $0.26 d_l$~\cite{wang2011prlb}. 

Near a $p$-wave resonance where the scattering volume $V_p$ goes through a pole, the scaling law of the three-body loss rate has been extracted numerically~\cite{wang2011prlb} as $L_3\propto k^4 V_p^{17/2}d_l^{35/2}$. Even though there is the $T^2$ suppression at ultracold temperatures, $L_3$ grows very quickly 
with $V_p$ and implies short-lived fermionic dipolar gases on resonance. Away from resonance, the loss is expected to have $d_l^8$ scaling and still 
implies reduced stability in the strong dipole regime. 

\section{Fewbody physics with other types of long-range interactions}\index{index!multiple indexes} 
\label{LongRange}

Long-range interactions bring in long-range correlations --- the craddle of universal physics. In few-body systems, long-range correlations 
often lead to binding of exotic quantum states. Such states have strong impact on the dynamics of atoms and molecules at ultracold temperatures, and provide 
convenient means in quantum controls in ultracold chemistry. An example is the the engineering of molecular states, as demonstrated by Innsbruck group to steer 
diatomics through high-lying 
ro-vibrational states using magnetic Feshbach resonances~\cite{mark2007pra,lang2008naturephys,zenesini2014pra}. 
Although the control over triatomics and poly-atomic molecular states is 
still in the beginning stage, microwave association of Efimov trimers has been experimentally achieved by~\cite{lompe2010science} and 
Machtey, {\it et al.}~\cite{machtey2012prl}. 
For future experimental development, it is important to understand the properties of exotic few-body states with other types of long-range interactions, as 
well as possible controls that are realizable. In this section we give a brief review over such systems. 

{\it Coulomb interactions}. Ultracold few-body systems with Coulomb force is very rich in exotic quantum physics. In a system of identical charges, although interesting 
phenomena such as ion Coulomb crystal~\cite{mavadia2013naturecomm} is expected in the many-body physics level, few-body physics is less intriguing 
because of the lack of binding. Systems 
with opposite charges, on the other hand, are much more complicated and have been one of the main subjects in studies of atoms and molecules for many decades. 
In such systems, the existence of Rydberg series in the two-body sub-systems and the emergence of quantum chaos near a three-body breakup threshold~\cite{xu2008pra} 
often bring difficulties in the quantum analysis of scaling laws. Such scaling laws are very important for atoms forming in cold plasma 
and has been more widely studied by semi-classical analyses 
~\cite{wannier1953pr,macek1995prl,macek1996pra} and classical Monte-Carlo simulations~\cite{mansbach1969pr,pohl2008prl} 

 {\it Attractive $1/r^2$ potential.} Depending on its strength, the attractive $1/r^2$ potential can either be supercritical with infinite number of two-body bound states or subcritical where no two-body state is bound. If we write the potential as
\begin{equation}
V(r)=-\frac{\alpha^2+1/4}{m r^2}\hbar^2
\end{equation}
for two identical bosons of mass $m$, the super and subcritical behaviors are determined by $\alpha^2>0$ and $\alpha^2<0$. In the supercritical case, as mentioned earlier 
the attractive $1/r^2$ is too singular at the origin to have a lower bound for the ground state energy 
--- a short-range cutoff is therefore necessary for a system to be physical. 
In the subcritical case, although the attractive $1/r^2$ singularity does not cause any ill behavior for two bosons, 
Ref.~\cite{guevara2012prl} has numerically showed that the three-boson system suffers a collapse 
in the ground state energy when $\alpha^2$ is the range $-0.0072\lesssim\alpha^2<0$. 
With a short-range cutoff, this collapse is avoided but the three-boson system is no less 
interesting --- it has an infinite number of bound states as a result of the long-range form of a numerically observed, effective hyperradial potential 
\begin{equation}
\label{Eq_Logpot}
U(R)\approx -\frac{\sqrt{\beta \ln(R/r_0)+\delta}}{2\mu_3 R^2}\hbar^2,
\end{equation}
with $\beta>0$ in the range of $\alpha^2$ where the non-regularized system collapses. The parameter $r_0$ is the characteristic length scale for 
the short-range cutoff. Both $\beta$ and $\delta$ depend on $\alpha^2$, 
whereas $\delta$ depends also on the detail of the short-range cutoff and is therefore non-universal~\cite{guevara2012prl}. 
The collapse of a non-regularized system can be understood by Eq.~(\ref{Eq_Logpot}) when the limit $r_0\rightarrow 0$ is taken --- $U(R)$ diverges for every $R$. 
So unlike the Thomas collapse, this collapse cannot be removed by simply introducing a short-range three-body force. 

For fermions, the supercritical behavior occurs when $\alpha^2>2$ due to their $p$-wave barrier. In the subcritical case, it is numerically found~\cite{guevara2012prl} 
that the effective hyperradial potential for three fermions has the asymptotic behavior
\begin{equation}
U(R)\approx -\frac{\alpha_{\rm eff}^2+1/4}{2 \mu_3 R^2}+\frac{\gamma}{2 \mu_3 R^2 \ln(R/r_0)},
\end{equation}
with $\alpha_{\rm eff}^2>0$ when $1.6\lesssim \alpha^2<2$. This again leads to a collapse of the three-body system without proper regularization in $V(r)$. 
Unlike the case of bosons, though, taking the $r_0\rightarrow 0$ limit here does not lead to a collapse in $U(R)$, 
therefore a three-body force at small $R$ is sufficient to avoid the collapse of the system. Here the bizarreness is that 
three fermions have an infinite number of bound states when $1.6\lesssim\alpha^2<1.75$, where 
the effective two-body interactions (including the $p$-wave barrier) are all repulsive! 

Although the attractive $1/r^2$ interaction doesn't exist between fundamental particles, similar type of interactions can be found between composite particles. 
One example is the interaction between a charge and dipole, although it is not obvious what changes the anisotropy could make in the three-body physics. 
Also, in such systems there can be at most two pairs of $1/r^2$ interactions, so that the results discussed above for identical particles do not apply directly. 
Another possible scenario is a system of three heavy and one light atoms with resonant heavy-light interactions (infinite heavy-light scattering length). In such systems 
the effective interaction between the heavy ones (induced by the light particle) is also attractive $1/r^2$ [see Eq.~(\ref{Eq_HHL})]. Here, though, the question is that if the 
way to get the effective interaction --- the Born-Oppenheimer approximation --- is good enough to study the new, exotic three-body states in the subcritical regime, 
where relatively weak effective interactions, or moderate mass ratios, are needed. 
Another caution is that the effective $1/r^2$ interactions in this system 
do not exist for all geometries of the heavy particles~\cite{amado1973prd}, where the possible effects are yet to be investigated. 

{\it Ion-atom interactions.} The attractive $1/r^4$ interaction between an ion and a neutral atom is of much longer range than the van der Waals interaction between neutral atoms. The direct consequence is that the density of state in the near-threshold two-body spectum for the ion-atom interaction is much 
higher that the van der Waals systems, and the characteristic length scale 
\begin{equation}
r_4=(2\mu_2 C_4)^{1/2}/\hbar
\end{equation}
for the $-C_4/r^4$ interaction is typically on the order of $10^3$ Bohr, which is much greater than $r_{\rm vdW}$ in atomic systems. 
Nevertheless, the near-threshold two-body physics for the 
attractive $1/r^4$ interaction is essentially the same as that for the van der Waals interaction~\cite{watanabe1980pra,gao2010prla,idziaszek2011njp,gao2013pra,li2014pra}, 
and so is expected for Efmov physics 
(assuming attractive $1/r^4$ interactions for all pairs)~\cite{naidon2014prl}. In realistic systems, however, inelastic collisions due to electron exchange 
are often non-negligible, which is beyond the physics that single-channel attractive $1/r^4$ interactions describe. 
Recent experimental studies~\cite{harter2012prl}, though, have suggested 
that in a system of one Rb$^+$ ion and two neutral Rb atoms the $Rb^+$ ion behaves more like a ``catalyst'' in the process of three-body recombination, and rarely appear 
in the recombination products. 
The three-body physics in the ion-neutral hybrid system, including electron exchange and the physics of long-range 
attractive $1/r^4$ interaction, is therefore an important topic for future studies to understand the dynamics in ion-neutral admixtures. 

{\it Spin-orbit interactions.} The term ``Spin-orbit interaction'', or ``Spin-orbit coupling'', is used both in the context of atomic physics and condensed matter physics. 
Here we refer to its latter meaning, namely, the interaction between a particle's spin or pseudo-spin $\boldsymbol{s}$ and its {\it linear} 
momentum $\boldsymbol{p}$ in the following form:
\begin{equation}
V=c  (\boldsymbol{s\cdot a}) (\boldsymbol{p\cdot b}). 
\end{equation}
where $\boldsymbol{a}$ and $\boldsymbol{b}$ are some constant vectors that depend on how the spin-orbit interaction is created~\cite{zhai2012intjmpb}. 
In condensed matter systems, the spin-orbit interactions play an important role in many exotic quantum phenomena, 
such as topological phases and quantum spin Hall effect~\cite{hasan2010rmp}. 
Thanks to the recent experimental realization of the spin-orbit interaction, or more generally, the synthetic gauge field in ultracold atomic gases (see a recent review ~\cite{galitski2013nature}), 
physics in spin-orbit coupled few-body systems has been a very active research subject in the last few years. 

Since particles with the spin-orbit interaction are no longer in their 
momentum eigenstates even when they are far apart, their low-energy scattering properties, as well as the density of states near the two-body breakup threshold 
differ dramatically from systems without spin-orbit interactions~\cite{vyasanakere2011prb}.  
New, exotic few-body physics is therefore highly expected in the spin-orbit coupled systems.  

Although still in the early stage, understanding in the two-body binding and proper treatments of two-body scattering have been 
developed for spin-orbit coupled systems 
with different identical particle symmetries, spin configurations, and various types of model 
potentials~\cite{cui2012pra,zhang2013pra,zhang2012pra,zhang2012pra2,wu2013pra,duan2013pra}. Such developments have 
facilitated many experimental observations, for instance, in the anisotropic low-energy scattering property~\cite{williams2012science} and in 
the Feshbach molecules creation controlled by spin-orbit interactions~\cite{fu2014naturephys}. 

Studies on three particles with spin-orbit interactions are just in the beginning. Nevertheless, universal three-body states induced by spin-orbit interactions have 
been predicted by~\cite{shi2014prl} 
in systems of mass imbalanced systems --- two heavy spinless fermions and one light, spin $1/2$ atom. In this 
system, extra binding comes from the lift of degeneracy by the spin-orbit interactions in states with the same total orbital angular momentum. 

\section{Summary}\label{Summary}\index{index!multiple indexes} 

Few-body physics, which connects fundamental quantum physics to macroscopic properties of many-body systems, serves as the fundation of modern quantum physics, 
quantum chemistry, molecular biology, and many other sciences and technologies. In comparison to larger systems, on the one hand, 
few-body systems are simple enough so that accurate 
descriptions and understandings can be obtained; on the other hand they are complex enough for many exotic quantum phenomena to occur 
and extensive experimental controls to 
be possible. 
In our above discussions we have briefly reviewed the universal properties of strongly interacting two- and three-body systems in ultracold atomic gases, 
where extraordinary 
level of control and tunability have been achieved over the interaction strength, spatial dimensions, identical particle symmetries, 
and many other configurations of few-body systems. Here in particular, studies on the van der Waals and dipolar atomic systems 
have indicated that a full characterization of three-body physics is in principle feasible from the relatively simple, yet accurately known two-body properties. 
Studies on the generalization of these new findings in other ultracold atomic systems, including those with experimentally synthesized interactions, are 
highly desirable and may lead to another milestone in the quantum few-body research. 

\section{Acknowledgments}

This work was supported in part by an AFOSR-MURI and by NSF. YW also acknowledges support from Department of Physics, Kansas State University.

\bibliographystyle{ws-rv-van}
\bibliography{ThreeBody.bib}
\blankpage
\printindex                         
\end{document}